\documentclass[hyper,11pt,paper]{article}
\usepackage{jheppub}
\usepackage[usenames,dvipsnames,table]{xcolor}
\usepackage{graphicx,amsmath,amssymb,multirow,array,bm,mathrsfs,bbold,bbm}
\usepackage{footmisc}
\usepackage{epsf,amsfonts,slashed,soul}
\usepackage{subfig}

\usepackage{pifont}

\usepackage{bbold}

\newcommand{\be}{\begin{equation}}
\newcommand{\ee}{\end{equation}}
\newcommand{\ba}{\begin{eqnarray}}
\newcommand{\ea}{\end{eqnarray}}
\newcommand{\nn}{\nonumber }

\title{Holographic Schwinger-Keldysh effective field theories}

\author[a]{Jan de Boer,}
\author[b*]{Michal P. Heller,\note[*]{On leave from: \emph{National Centre for Nuclear Research, 00-681 Warsaw, Poland}.}}
\author[c]{and Natalia Pinzani-Fokeeva}
\affiliation[a]{Institute for Theoretical Physics, University of Amsterdam,
1090~GL Amsterdam, The Netherlands}
\affiliation[b]{Max Planck Institute for Gravitational Physics, Potsdam-Golm, D-14476, Germany}
\affiliation[c]{Institute for Theoretical Physics, KU Leuven,
 Celestijnenlaan 200D, Leuven B-3001, Belgium}

\emailAdd{j.deboer@uva.nl}
\emailAdd{michal.p.heller@aei.mpg.de}
\emailAdd{natascia.pinzanifokeeva@kuleuven.be}
\abstract{We construct a holographic dual of the Schwinger-Keldysh effective action for the dissipative low-energy dynamics of relativistic charged matter at strong coupling in a fixed thermal background. To do so, we use a mixed signature bulk spacetime whereby an eternal asymptotically anti-de Sitter black hole is glued to its Euclidean counterpart along an initial time slice in a way to match the desired double-time contour of the dual field theory.  Our results are consistent with existing literature and can be regarded as a fully-ab initio derivation of a Schwinger-Keldysh effective action. In addition, we provide a simple infrared effective action for the near horizon region that drives all the dissipation and can be viewed as an alternative to the membrane paradigm approximation.
}
\begin{document}
\maketitle


\section{Introduction}
\label{S:intro}

The path integral formulation of quantum mechanics is central to many among the most important developments in modern theoretical physics. For example, it played a pivotal role in the quantization of gauge theories. The latter are the corner stone of the Standard Model of particle physics~\cite{tHooft:2005hbu}, provide, through holography (AdS/CFT)~\cite{Maldacena:1997re,Witten:1998qj,Gubser:1998bc}, our current best shot at understanding quantum gravity and find numerous applications in the condensed matter physics~\cite{shankar_2017}.

Path integrals are also widely used to describe non-equilibrium phenomena. In these cases they are defined on nontrivial time contours in which time can evolve forward and backwards in various sequences. A special case of this type is the Schwinger-Keldysh partition function~\cite{Schwinger:1960qe,Keldysh:1964ud}. This object incorporates a particular class of correlators which involve time-ordering (i.e. retarded, advanced, symmetric, etc.) and has been extensively used in numerous applications to non-equilibrium collective states of matter both in high energy and condensed matter physics, see, e.g., \cite{2004cond.mat.12296K} and references therein. 
 
One of the most active lines of research on non-equilibrium phenomena of the past two decades concerns relativistic hydrodynamics. The key experimental impetus on this front was the discovery that  droplets of quark-gluon plasma created in ultrarelativistic heavy-ion collisions at RHIC (and later also at LHC) behave as the most perfect fluid known in the universe. It was also understood relatively early on that deviations from perfect fluidity provide an invaluable window on the microscopics of the quark-gluon plasma. See,~e.g.,~ref.~\cite{Busza:2018rrf} for a recent broad overview of this research field. Altogether, this has led to a large number of developments concerning dissipative relativistic hydrodynamics that have been occurring both on the phenomenological and theoretical front, see, e.g., refs.~\cite{Romatschke:2009im,Hubeny:2011hd,Florkowski:2017olj,Romatschke:2017ejr} for a review. 

The vast majority of the aforementioned progress occurred at the level of equations of motion, which has limited our ability to understand the microscopic origin of dissipation and the entropy current, as well as to systematically incorporate noise. It was only very recently that the  Schwinger-Keldysh formulation has been used to bring us closer to recasting the relativistic fluid dynamics as a proper effective field theory with a Lagrangian formulation, see refs.~\cite{Haehl:2015foa,Crossley:2015evo,Haehl:2015uoc,Haehl:2016pec,Haehl:2016uah,Glorioso:2017lcn,Jensen:2017kzi,Gao:2017bqf,Glorioso:2017fpd,Haehl:2018lcu,Jensen:2018hse}, the precursor works~\cite{Grozdanov:2013dba,Kovtun:2014hpa,Haehl:2014zda,Haehl:2015pja,Harder:2015nxa} as well as ref.~\cite{Glorioso:2018wxw} for a review and a more comprehensive list of references. 

The symmetries of the microscopic Schwinger-Keldysh path integral, such as unitarity and CPT invariance, have been used to constrain the form of the low-energy Schwinger-Keldysh effective action and consequently to re-derive the conventional phenomenological formulation of hydrodynamics in some limiting regimes. For example, it was shown in ref.~\cite{Glorioso:2016gsa}, see also~refs.~\cite{Jensen:2018hhx, Haehl:2018uqv}, how the local second law of thermodynamics can be obtained as the Noether current of a symmetry of the effective action. Moreover, additional constraints on transport, invisible to the phenomenological treatment based on equations of motion, have been found in~ref.~\cite{Jensen:2018hse}. As alluded earlier, the power of this approach is that both dissipative and fluctuation effects are naturally incorporated allowing for a systematic treatment of stochastic noises, see the recent ref.~\cite{Chen-Lin:2018kfl} which evaluates corrections to the heat diffusion coefficient due to thermal fluctuations finding novel results with respect to previous, more phenomenological, derivations, see, e.g., ref. \cite{Kovtun:2014nsa}.

The key motivation for our present work is that the above considerations, while agreeing with all previously known results that stood the test of scrutiny and addressing some of their shortcomings, have not yet been derived from any microscopic model when dissipation needed to be included. Our aim, therefore, is to provide the first microscopic derivation of the Schwinger-Keldysh effective actions for what seems to be the simplest hydrodynamic setup, i.e. the charge diffusion in the limit in which the energy and momentum carried by charge carriers is negligible and the charge current amplitude is small.

The tool that we will use to achieve this goal is holography viewed here as an \emph{ab initio} formulation of a large class of strongly-interacting quantum field theories with a large number of microscopic constituents. 

Indeed, holography provides a very natural arena in which all these ideas can be explicitly tested. Following ref.~\cite{Herzog:2002pc}, the double-time contour of the Schwinger-Keldysh formalism can be represented by the two boundaries of the maximally-extended AdS-Schwarzschild black brane. Subsequent developments in refs.~\cite{Skenderis:2008dh,Skenderis:2008dg} interpreted the Euclidean part of the thermal Schwinger-Keldysh contour as a
Euclidean black brane glued to a finite time slice of the Lorentzian solution. The contour can be then closed by gluing the future horizons of the Lorentzian solution as in ref.~\cite{vanRees:2009rw}, see also ref.~\cite{Botta-Cantcheff:2018brv}. This simple picture, of course, applies only if the sources and their effects are a tiny distortion of the global thermal equilibrium. Otherwise, according to the fully holographic prescription of refs.~\cite{Skenderis:2008dh,Skenderis:2008dg}, one would need to solve non-linear Einstein's equations (possibly with matter) and find a mixed signature holographic geometry specified by the asymptotic boundary condition given by the desired Schwinger-Keldysh contour. In the present work we circumnavigate this otherwise fascinating problem by considering the dynamics of a probe bulk gauge field  in the aforementioned mixed-signature static background spacetime. This scenario gives then rise to the quadratic Schwinger-Keldysh effective action for the charge diffusion in holographic conformal field theories, which we derive to the second order in derivatives.

Our work builds on several earlier developments. In particular, in the previous paper~\cite{deBoer:2015ija}, see also ref.~\cite{Crossley:2015tka}, we considered the preliminary case of dissipationless fluids, i.e. those which do not exhibit entropy production, and derived their leading order effective action from holography. Our approach then focused on single-sided, asymptotically AdS black branes dual to a strongly-coupled conformal plasma around thermal equilibrium. To remove dissipative effects, we introduced a fictitious intermediate timelike hypersurface at some fixed radial position very close to the horizon and considered only the remaining spacetime between this radial cut-off and the conformal boundary. We identified the holographic dual of the   degrees of freedom relevant for the fluid dynamic behavior with Wilson lines extending between the conformal boundary and the cutoff hypersurface, much in the spirit of~ref.~\cite{Nickel:2010pr}, see also refs.~\cite{Heemskerk:2010hk,Faulkner:2010jy}. After solving the linearized Einstein equations with double-Dirichlet boundary conditions between the two boundaries at the leading order in the hydrodynamic gradient expansion, we computed the (partially) on-shell gravitational action which we then interpreted as the effective action for ideal fluids. It was, of course, already clear back then that this procedure was not enough had one wished to go beyond the leading order hydrodynamic expansion. Not only the dissipation was absent by construction, but also non-dissipative second order transport was suffering from divergences. The latter could only be cured by including dissipative effects, as we showed there using a simple membrane paradigm approximation, see in this context also our earlier work~\cite{deBoer:2014xja}.

The plan of the present article is the following. In sec.~\ref{S:SK} we summarize the key ingredients of the Schwinger-Keldysh formalism which are necessary for the present work. The reader familiar with the literature of the subject may skip this part. In sec.~\ref{S:hol} we provide a general discussion on how to parallel the same construction in holography using the real-time formulation of refs.~\cite{Skenderis:2008dh,Skenderis:2008dg}. In sec.~\ref{S:diff} we  consider the specific example of a probe gauge field and show how the low-energy effective action for charge dynamics at strong coupling can be \emph{derived} from holography. In sec.~\ref{S:IRmaster} we show how the general properties of the Schwinger-Keldysh effective action are captured by a simple infrared (IR) piece which resides in the near horizon region of the spacetime and how the remaining geometry between the horizon and the boundaries only serves as a map of this information to where the dual field theory is defined. Finally, in sec.~\ref{S:Discussion} we discuss some natural extensions of our results.

\vspace{10 pt}

\noindent{\bf Note added:} We learnt that ref.~\cite{Glorioso:2018mmw}, which will appear soon, also addresses the problem of deriving Schwinger-Keldysh effective actions using holography.

\noindent{\bf Note added (v2):} The results of ref.~\cite{Glorioso:2018mmw} are in agreement with ours.

\section{The Schwinger-Keldysh effective action}
\label{S:SK}

Let us briefly review the main ingredients that enter the construction of  Schwinger-Keldysh effective actions. We will mainly focus on  the simplest case of the low-energy behavior of charged matter in a fixed thermal background which will be used in this work. This section is admittedly brief, we refer  the reader to the  review \cite{Glorioso:2018wxw} and references therein for more details.

\begin{figure}[t]
\centering
 \includegraphics[width=5cm]{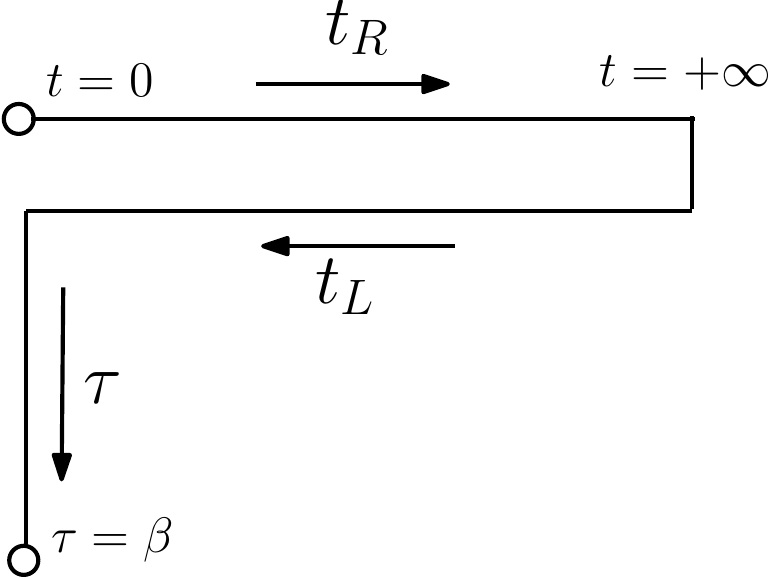}
\caption{The Schwinger-Keldysh contour at finite temperature. Time flows forward from $t=0$ to $t=+\infty$ and back to where the initial state is defined. The latter is represented by an imaginary time segment identified along the circles with period $\beta=1/T$.}
\label{fig.closed}
 \end{figure}

The Schwinger-Keldysh partition function is defined as
\be\label{SK:def}
Z[A_R,A_L]={\rm Tr}\Big({\cal U}[A_R]\,\rho_0\, {\cal U}^\dagger[A_L]\Big)\,,
\ee
where 
$\rho_0$ is the density matrix of the initial state given at some time $t=0$, ${\cal U}$ is the evolution operator from $t=0$ to the infinite future $t=+\infty$, and ${\cal U}^\dagger$ is the anti-evolution operator from the infinite future back to $t=0$ where the state is defined. The evolution operators depend generically on external sources,   collectively denoted by $A_R$ and $A_L$, which are taken to be independent from one another. In a path integral representation, eq.~\eqref{SK:def} takes the form
\be\label{SK:def2}
Z[A_R,A_L]=\int_{\rho_0} {\cal D}\psi_R \, {\cal D}\psi_L \, e^{iS[\psi_R;\,A_R]-iS[\psi_L;\,A_L]}
\,,
\ee
where $S$ is the microscopic action, functional of the microscopic fields $\psi$ and the sources $A$. Appropriate boundary conditions need to be supplemented  in the infinite future $t=+\infty$, i.e. $\psi_L(t=+\infty)=\psi_R(t=+\infty)$, and at $t=0$. The relative minus sign between the $R$ and $L$-type actions in \eqref{SK:def2} reflects the backward flow of time of the anti-evolution operator in eq.~\eqref{SK:def}. With these definitions, the Schwinger-Keldysh path integral~\eqref{SK:def2} can be viewed as an integral over a double-time contour where time flows forward in the upper part of the contour with time $t_R$ and then back with  time $t_L$ to where the state is defined. If the initial state is thermal, i.e. $\rho_0\sim e^{-\beta H}$ with  the inverse temperature $\beta=1/T$, then the contour becomes complex, as depicted in~Fig.~\ref{fig.closed}, and the imaginary time segment is also periodically identified with period~$\beta$.

 The main virtue of the Schwinger-Keldysh partition function \eqref{SK:def2} is that it incorporates a large class of Minkowski signature correlators for generic states. Defining the Schwinger-Keldysh average and difference basis
 \be
 A_r=\frac{1}{2}\left(A_R+A_L\right)\,,\qquad A_a=A_R-A_L\,,
 \ee
 the retarded, advanced and symmetric connected two-point functions  can be shown to be related to \emph{very simple} (which is part of the formalism's allure) variations of the generating functional $W=-i\ln Z$ as follows
 \begin{align}
  \begin{split}\label{eq:Green}
 G_{ret}(t_1,t_2) & =  i\,G_{ra}(t_1,t_2) =i \frac{\delta^2 W}{\delta A_{a}(t_1)\delta A_{r}(t_2)}\bigg|_{A=0} \,,\\
  G_{adv}(t_1,t_2) & =  i\,G_{ar}(t_1,t_2) =i \frac{\delta^2 W}{\delta A_{r}(t_1)\delta A_{a}(t_2)}\bigg|_{A=0} \,,\\
   G_{sym}(t_1,t_2) & =  G_{rr}(t_1,t_2) = \frac{\delta^2 W}{\delta A_{a}(t_1)\delta A_{a}(t_2)}\bigg|_{A=0} \,,
\end{split}
 \end{align}
 see \cite{Glorioso:2018wxw} for more details. These functions are of obvious interest in many-body physics\footnote{In particular, the structure of singularities in the complex frequency and momentum plane of retarded two-point functions has been the driving force behind recent progress on understanding non-equilibrium properties of states of matter evading quasiparticle description, see, e.g., refs.~\cite{Hartnoll:2016apf,Florkowski:2017olj,Liu:2018crr} for contemporary reviews.}.

Let us now separate in eq.~\eqref{SK:def2} between the high-energy (UV) and low-energy (IR) degrees of freedom. Assuming that the UV degrees of freedom have been integrated out, the Schwinger-Keldysh path integral~\eqref{SK:def2} can be formally rewritten 
 in terms of appropriately defined IR dynamical fields $\xi_R$ and $\xi_L$ as
\be\label{SK:eff}
Z[A_R,A_L]=\int {\cal D}\xi_R \, {\cal D}\xi_L \, e^{i\,S_{eff}[\xi_R,\,\xi_L;\,A_R,\,A_L;\,\rho_0]}\,,
\ee
where $S_{eff}$ is the Schwinger-Keldysh effective action that incorporates the  low-energy dynamics around the state $\rho_0$.
With respect to the conventional Wilsonian effective action  there are a few differences. For example, the Schwinger-Keldysh effective action depends on two copies of the infrared degrees of freedom which  may interact with one another. It is those interactions that allow to write  dissipative terms in the Lagrangian, something that is impossible to achieve with  conventional effective actions.  

In this work  we are interested in the low-energy effective action $S_{eff}$ for  charged matter at finite temperature in a fixed  $d$-dimensional   background. In this case, the external  sources are taken to be a pair of $U(1)$ flavor fields $A_{R\,\mu}$ and $A_{L\,\mu}$ which live on two different spacetimes, dubbed as $R$ and $L$, and transform as connections under two independent gauge transformations parameterized by $\Lambda_R$ and $\Lambda_L$,
\be
A_{R\,\mu}\rightarrow A_{R\,\mu}+\partial_{\mu} \Lambda_R\,,\qquad A_{L\,\mu}\rightarrow A_{L\,\mu}+\partial_{\mu} \Lambda_L\,,
\ee
where $\mu=t\,,x_1\,,\dots\,, x_d$ are spacetime indices. Notice that the coordinates defined on the   $R$ and $L$ spacetimes, $x^{\mu}=(t,\vec{x})$, can in principle be independent as the fields in $R$ transform under independent diffeomorphisms than those in $L$. However,   as we already alluded many times, we are only interested in the charge dynamics, thus in a probe limit where the background  is fixed to the Minkowski metric $\eta_{\mu\nu}$ and the spacetime coordinates on which the two type of fields $R$ and $L$ depend can be taken to be  identical.   

The relevant low-energy degrees of freedom  are two scalar fields $\phi_L$ and $\phi_R$ that appear in the effective action in  combination  with the sources as follows
\be\label{eq:B}
B_{R\,i}(\sigma)=\partial_ix^{\mu}(\sigma)A_{R\,\mu}(x(\sigma))+\partial_{i}\phi_R(\sigma)\,,\qquad
B_{L\,i}(\sigma)=\partial_ix^{\mu}(\sigma)A_{L\,\mu}(x(\sigma))+\partial_{i}\phi_L(\sigma)\,.
\ee
Expressions \eqref{eq:B} are pullbacks   of the sources $A_{R\,\mu}$ and $A_{L\,\mu}$ from their respective $R$ and $L$ {\it target} spaces to a common  {\it worldvolume} spacetime parameterized  by the coordinates $\sigma^i=(\bar{t},\vec{x})$. The maps $x^{\mu}(\sigma)$, $\phi_R(\sigma)$ and $\phi_L(\sigma)$ define the pullback and here we have promoted the scalar fields to be the dynamical low-energy fields that capture the hydrodynamic behavior of the charge current. The remaining maps $x^{\mu}(\sigma)$ are not dynamical and are taken to only include a reparameterization of time between the target spaces and the worldvolume. With a little bit of prescience we  write them as   $x^{\mu}(\sigma)=\left(\bar{t}/\sqrt{f_{\delta}},\vec{x}\right)$ where $\sqrt{f_{\delta}}$ is a redshift factor. As we will see again later on, this picture is very much reminiscent of  black holes where the target spaces can be identified with the boundaries of an eternal AdS black hole and the worldvolume with the near horizon region where the redshift can be naturally defined through the emblackening factor. In fact, the metric on the worldvolume is defined as 
\be\label{eq:worldmetric}
g_{ij}(\sigma)=\partial_i x^{\mu}(\sigma)\partial_j x^{\nu}(\sigma)\eta_{\mu\nu}(x(\sigma))\,,
\ee 
which is the pullback of the boundary metric to the near horizon region.

The scalar fields $\phi_R$ and $\phi_L$ are  respectively bifundamental maps between the tangent bundles over the $R/L$ target spaces and the tangent bundle over the worldvolume. They transform thus as follows
\be\label{SK:bi}
\phi_{R}\rightarrow \phi_R-\Lambda_R+\Lambda\,,\qquad \phi_L\rightarrow \phi_L -\Lambda_L + \Lambda\,,
\ee
where $\Lambda$ is the $U(1)$ gauge parameter on the worldvolume. With these definitions, the combinations \eqref{eq:B} are invariant under the  $R$ and $L$ gauge transformations, and transform as connections under the worldvolume one
\be
B_{R\,i}\rightarrow B_{R\,i}+\partial_{i} \Lambda\,,\qquad B_{L\,i}\rightarrow B_{L\,i}+\partial_{i} \Lambda\,.
\ee
In what follows, it is convenient to introduce $r$ and $a$-type combinations of the fields and sources 
\be\label{eq:redar}
\phi_r=\frac{1}{2}\left(\phi_R+\phi_L\right)\,,\quad \phi_a=\phi_R-\phi_L\,,\quad B_{r\,i}=\frac{1}{2}\left(B_{R\,i}+B_{L\,i}\right)\,,\quad B_{a\,i}=B_{R\,i}-B_{L\,i}\,.
\ee

Having defined the low-energy degrees of freedom, the most general Schwinger-Keldysh effective action $S_{eff}$ can now be constructed requiring compatibility with the symmetries of the Schwinger-Keldysh path integral \eqref{SK:def}, 
\ba
\text{Schwinger-Keldysh symmetry:}&& \quad Z[A_{R}=A_L]=1\,,\label{eq:SKsymm}
\\
\text{Reality condition:}&&\quad Z[A_R,A_L]^*=Z[A_L^*,A_R^*]\,,\label{eq:SKreality}
\\
\text{KMS symmetry:}&&\quad  Z[A_R(t),A_L(t)]= Z[\eta A_R(-t),\eta A_L(-t-i\beta)]\label{eq:KMSsymm}
\,,\qquad 
\ea
 and with the constraint
 \be\label{cond4}
\text{Im}\,S_{eff}\geq 0\,.
\ee
 Expressions \eqref{eq:SKsymm} and \eqref{eq:SKreality}  are valid for any initial state $\rho_0$ while the KMS symmetry \eqref{eq:KMSsymm} is only applicable for thermal states $\rho_0\sim e^{-\beta{\cal H}}$ where $\beta=1/T$ is the inverse of the temperature,
see, again, ref.~\cite{Glorioso:2018wxw} for more details. In writing eq.~\eqref{eq:KMSsymm} we have also assumed CPT invariance\footnote{The condition on CPT invariance of the microscopic theory can be changed to any combination of the discrete symmetries as long as it contains time invariance, see ref.~\cite{Crossley:2015evo}.} of the underlying microscopic theory, where $\eta$ is the CPT eigenvalue associated to the operator dual to the corresponding source $A$. For example, in the case $A$ is an external $U(1)$ flavor field $A_{\mu}$ depending  on $t$ and $\vec{x}$ coordinates of the Minkowski metric, the corresponding CPT eigenvalue $\eta_{\mu}$  is $\eta_t=\eta_x=-1$ and the remaining components are $+1$.

Notice that upon reinstating $\hbar$ in the expressions above we have  $\beta\rightarrow \hbar \, \beta$, and so the KMS symmetry~\eqref{eq:KMSsymm} is a nonlocal $Z_2$ symmetry at the full quantum level.  Although it is possible to take a classical limit $\hbar\rightarrow 0$ where quantum fluctuations are neglected and the condition~\eqref{eq:KMSsymm} becomes local, for the probe limit case at hand this restriction is not necessary and we may work in the full quantum regime, see, e.g., ref.~\cite{Jensen:2017kzi}. 
Let us also notice that when $A_R(t)=A_L(t-i\beta)$, we have from \eqref{eq:KMSsymm} that $Z[A_R(t)=A_L(t-i\beta)]=1$, meaning that  the Schwinger-Keldysh partition function for thermal states has an additional limit where it becomes trivial, other than the one given in eq.~\eqref{eq:SKsymm}.

We cannot help but mention that the Schwinger-Keldysh symmetry \eqref{eq:SKsymm} can be implemented at tree level by requiring the effective action to be at least linear in the $a$-type combinations of the fields and sources $B_a$.
At the  full quantum level however this condition is not sufficient. The same property can be accounted for using a  BRST-type symmetry introducing additional ghost-like degrees of freedom~\cite{Haehl:2015foa,Crossley:2015evo,Haehl:2015uoc,Haehl:2016pec,Haehl:2016uah}, see also refs.~\cite{Haehl:2015uoc,Haehl:2016pec,Haehl:2016uah,Jensen:2017kzi,Gao:2017bqf,Haehl:2018lcu,Jensen:2018hse} for a supersymmetric implementation of this symmetry\footnote{A second BRST charge arises when imposing the KMS symmetry \eqref{eq:KMSsymm}, see refs.~\cite{Crossley:2015evo,Jensen:2017kzi,Gao:2017bqf,Jensen:2018hse}. The authors of refs.~\cite{Haehl:2015foa,Haehl:2015uoc,Haehl:2016pec,Haehl:2016uah} have a more general structure of an ${\cal N}=2$ equivariant cohomology.}. While such a construction is  convenient as an organizational principle, it has been shown in ref.~\cite{Gao:2018bxz} that ghost degrees of freedom effectively decouple to all loop orders in the effective action and, for the purposes of this work, we will neglect them completely. 

Another set of symmetries of the Schwinger-Keldysh effective action comes from the doubled structure of the path integral \eqref{SK:def2}: every continuous symmetry of the microscopic action is doubled.   In our case, this amounts to have a  doubled  $U(1)$ flavor symmetry which can be implemented at the level of the Schwinger-Keldysh effective action by allowing the low-energy degrees of freedom and the external sources to appear only in the 
  invariant pullback combinations \eqref{eq:B},
  \be\label{eq:action}
  S_{eff}=S_{eff}[B_{r\,i},B_{a\,i}].
  \ee  
  
 The effective action \eqref{eq:action} not only needs to be invariant under the doubled symmetries of the $R/L$ target spaces but also under the $U(1)$ symmetry implied on the common worldvolume spacetime by the  transformation properties of the bifundamental fields \eqref{SK:bi}. This transformation is however restricted to preserve the  properties of the initial state. For example, in the coordinates where the chemical potential is given by the time component of the average pullback source $\mu = B_{r\,\bar{t}}$,   only time-independent gauge transformations with parameter $\Lambda(\vec{x})$ are allowed to keep   $\mu$   invariant. Thus, under the ``gauge-fixed'' worldvolume $U(1)$ gauge transformations, the pullback sources transform~as\footnote{ The transformations \eqref{cond5} are given in the so-called static gauge. See, e.g., refs.~\cite{Haehl:2015pja,Jensen:2017kzi} for a full covariant implementation of the worldvolume gauge invariance.}
\be\label{cond5}
B_{r\,i}\rightarrow B_{r\,i} + \partial_{i} \Lambda(\vec{x})\,, \qquad B_{a\,i}
\rightarrow B_{a\,i}\,,
\ee
and the effective action $S_{eff}$ in \eqref{eq:action} must be invariant under these transformations.

Having introduced all the symmetries, a local low-energy Schwinger-Keldysh effective action  can be written  in a  hydrodynamic expansion 
\be\label{SK:exp}
S_{eff} = S^{(0)}_{eff}+S^{(1)}_{eff}+\dots
\ee
where the superscript counts the amount of derivatives. The superficial\footnote{One of the most interesting recent lessons about hydrodynamics is that constitutive relations for the energy-momentum tensor truncated at low orders of the derivative expansion can work remarkably well even if the relevant gradient terms are of the order of the perfect fluid contributions. The naive criterion~\eqref{eq.lmfp} is then strongly violated. This, however, does not lead to a contradiction since it has been explicitly shown in several cases that the gradient expansion of the energy-momentum tensor is a divergent series for which convergence conditions appropriate for series with a finite radius of convergence do not apply. See ref.~\cite{Florkowski:2017olj} for a broad overview of the relevant developments.} dimensionless expansion parameter is
\be
\label{eq.lmfp}
\frac{l_{mfp}}{l}\ll 1,
\ee  
where $l$ is the size of the gradients and, in weakly-coupled theories, $l_{mfp}$ is the typical scale for consequent collisions of the microscopic particles. For conformal field theories, $l_{mfp}\sim\hbar/T$ with the proportionality constant in general depending on the interaction strength\footnote{For holographic theories, for which the coupling constant is infinite, the proportionality constant turns out to be just some finite number.}, see,~e.g.,~ref.~\cite{Rangamani:2009xk}. 

To count derivatives, we choose $B_{r}\sim{\cal O}(1)$ and $B_{a}\sim {\cal O}(\partial)$, since only in this way terms related by the KMS symmetry \eqref{eq:KMSsymm} appear on the same footing. This can be seen by realizing that $A_L(t-i\beta)\sim A_L(t)-i\,\beta \, \partial_t A_{L}(t)+{\cal O}(\partial^2)$ and that in the same limit the average and difference combinations transform as $A_r\rightarrow A_r$ and $A_a \rightarrow A_a+i\beta \partial_t A_r$ under the KMS symmetry \eqref{eq:KMSsymm}. Thus, for example, a term in the effective action of the form $B_a\,\partial B_r $ is related to a term proportional to $B_a B_a$. 

 Finally, the physical equation of motion for the dynamical field $\phi_r$ is associated to the variation of the effective action with respect to the  field $\phi_a$\footnote{There is an equivalent equation  when varying the effective action with respect to the $\phi_r$ fields. In the absence of difference sources $A_a=0$ it is solved by setting $\phi_a=0$.} and it can be  recast into a conservation equation 
\begin{equation}
\frac{\delta S_{eff}}{\delta \phi_a}\bigg|_{B_a=0}=-\partial_{i}J^{i}=0 \,,
\end{equation}
where  the charge current is defined as
\begin{equation}\label{eq:const}
J^{i} = \frac{\delta S_{eff}}{\delta B_{a\,i}}\bigg|_{B_a=0}\,.
\end{equation}
In this way it is possible to extract the off-shell ($\phi_r$-dependent, i.e. obtained without solving the equations of motion for $\phi_{r}$) constitutive relations  for the charge current $J^{i}$ from an effective action $S_{eff}$ order by order in a derivative expansion. Having $J^i$,  the constitutive relations in the target space(s) can be simply obtained via a pushforward
\be\label{eq:push}
J^{\mu}=\partial_i x^{\mu} J^i\,.
\ee
Given that we will be almost exclusively  interested in quantities defined on the target spaces and that the pullback maps for the $U(1)$ flavor field in the probe limit are rather simple, in what follows we will construct the effective action directly in the target space(s).
 
Combining all the ingredients together, the most general local Schwinger-Keldysh effective action describing the dynamics of a conserved charge current  can be constructed using $B_r$ and $B_a$ fields satisfying the conditions \eqref{eq:SKsymm}, \eqref{eq:SKreality}, \eqref{eq:KMSsymm} and \eqref{cond5}. To   third order in a derivative expansion,\footnote{ Note that we assigned $B_a\sim {\cal O}(\partial)$ and since the effective action is at least linear in $B_{a}$, keeping terms up to third order in derivatives in the effective action amounts to having the constitutive relations up to  second order in derivatives.} to quadratic order in the fields and restricting to the longitudinal sector  where the wave vector points along the $x$ direction $\vec{k}=(k,0,\dots)$ and the only non-vanishing space component of $B_i$ is $B_x$, the most general effective action (see Appendix \ref{A:first} for  details)  is given by
\begin{align}
\begin{split}
\label{eq:Seff}
S_{eff}&=\int d^dx\Big( \chi\, B_{r\,t}B_{a\,t}-\sigma\,   B_{a\,x}\partial_{t}B_{r\,x}+i\,\sigma\, T\,   B_{a\,x}B_{a\,x}-s\,B_{a\,t}\partial_tB_{r\,t}+i\,s\,T\,B_{a\,t}B_{a\,t}\\
&\qquad+k_0B_{a\,t}\partial_t^2B_{r\,t}+k_1\, B_{a\,t}\partial_x^2 B_{r\,t} +k_2\,(\partial_xB_{a\,t}\partial_{t} B_{r\,x}+\partial_{t}B_{a\,x}\partial_xB_{r\,t} )+k_3\,B_{a\,x}\partial^2_{t}B_{r\,x}\Big)\,.
\end{split}
\end{align}
Here $\chi,\,\sigma,\,\dots$ are generic functions of the temperature $T$ which would need to be determined from the microscopic theory. Notice that the effective action \eqref{eq:Seff} is linear in $B_a$ such that the Schwinger-Keldysh symmetry \eqref{eq:SKsymm} is satisfied at least at tree level and the quadratic term in the $B_a$ fields is imaginary so that the reality condition \eqref{eq:SKreality} holds.  Moreover, the second and the third term (as well as the fourth and the fifth, and the one proportional to  $k_2$) in \eqref{eq:Seff}  are not independent since they are related to one another by   the KMS symmetry \eqref{eq:KMSsymm}. Finally, for \eqref{cond5} to be satisfied, the spatial components of $B_r$ fields only appear with time derivatives~$\partial_tB_{r\,x}$\,.

Finally, integrating out the dynamical degrees of freedom $\phi_a$ and $\phi_r$ in eq.~\eqref{eq:Seff} leads to the tree level generating functional $W=-i\ln Z$ from which one can extract the two-point functions~\eqref{eq:Green}. The diffusive pole can be computed from the retarded correlator and it is given by 
\be\label{eq:diffsecond}
\omega= - iD k^2 -i D k^4\left(k_1-k_2+k_3-sD\right)/\chi\,,
\ee
where we can readily interpret
\be
D=\frac{\sigma}{\chi}
\ee
as the diffusion constant, $\sigma$ as the conductivity and $\chi$ as the susceptibility. Note that it is a linear combination of several coefficients from the action that contributes to the second order (quartic in momentum) correction to the dispersion relation~\eqref{eq:diffsecond}. To close this part of the analysis, let us also comment that eq.~\eqref{cond4} constrains the first order coefficients to be non-negative
\be
\sigma\geq0\,,\qquad s\geq0\,,
\ee
see, e.g., refs.~\cite{Glorioso:2016gsa,Jensen:2017kzi}.

\section{Towards the holographic Schwinger-Keldysh effective action}
\label{S:hol}

In this section we build a gravitational parallel with the various field theory notions introduced so far. The main goal is to give a prescription for constructing and deriving the holographic dual of the low-energy Schwinger-Keldysh effective action $S_{eff}$ for strongly-coupled conformal field theories at finite temperature and a large rank of the gauge group~$N$ (large central charge).

\subsection{The mixed signature bulk spacetime}

A simple prescription to compute the Schwinger-Keldysh partition function \eqref{SK:def2} in holography appeared first in 
ref.~\cite{Herzog:2002pc} where the double-time contour of the dual field theory  was mimicked  in the gravity side by  an eternal black hole in AdS. The key ingredient of this construction is  the presence of an independent dual conformal field theory at each of the (left and right)  boundaries, see, e.g., refs.~\cite{1976PhLA...57..107I,Balasubramanian:1998de,Horowitz:1998xk,CarneirodaCunha:2001jf,Maldacena:2001kr}. In this way, a map between the  field theory Schwinger-Keldysh path integral \eqref{SK:def2} and the gravitational one  arises quite naturally by associating the two boundary field theories  to the doubled structure in eq.~\eqref{SK:def2}.

 The prescription of ref.~\cite{Herzog:2002pc} was subsequently refined in refs.~\cite{Skenderis:2008dh,Skenderis:2008dg} which provided a  systematic approach to real-time holography, previous works in this context include also refs.~\cite{Balasubramanian:1998sn,Son:2002sd,Satoh:2002bc,Kraus:2002iv,Marolf:2004fy,Lawrence:2006ze}. For example, while ref.~\cite{Herzog:2002pc} relied on certain {\it natural boundary conditions} to be imposed in the interior of the spacetime, the authors of refs.~\cite{Skenderis:2008dh,Skenderis:2008dg} adopted a fully  holographic point of view  where bulk dynamics  depends only on boundary data and at most on regularity conditions in the interior. Moreover, the procedure outlined in refs.~\cite{Skenderis:2008dh,Skenderis:2008dg} is not restricted to Schwinger-Keldysh contours but encompasses multiple forward and backward evolutions starting from a  general, not necessarily thermal, initial state. This powerful procedure can be summarized as follows. First, to each real-time segment of the contour associate a Lorentzian bulk spacetime and to each imaginary time segment assign a Euclidean one. All parts of the spacetime should  then be glued smoothly together along fixed time slices and any other field living in such mixed signature spacetime must also be smooth across the separation surfaces. Notice that the resulting mixed signature bulk spacetimes are not exotic and that similar constructions have been considered previously in the context of cosmology, see ref.~\cite{PhysRevD.28.2960} and a discussion of several related developments in footnote~\ref{footnote:complexmetrics}.

\begin{figure}%
    \centering
    \subfloat[]{{\includegraphics[width=5cm]{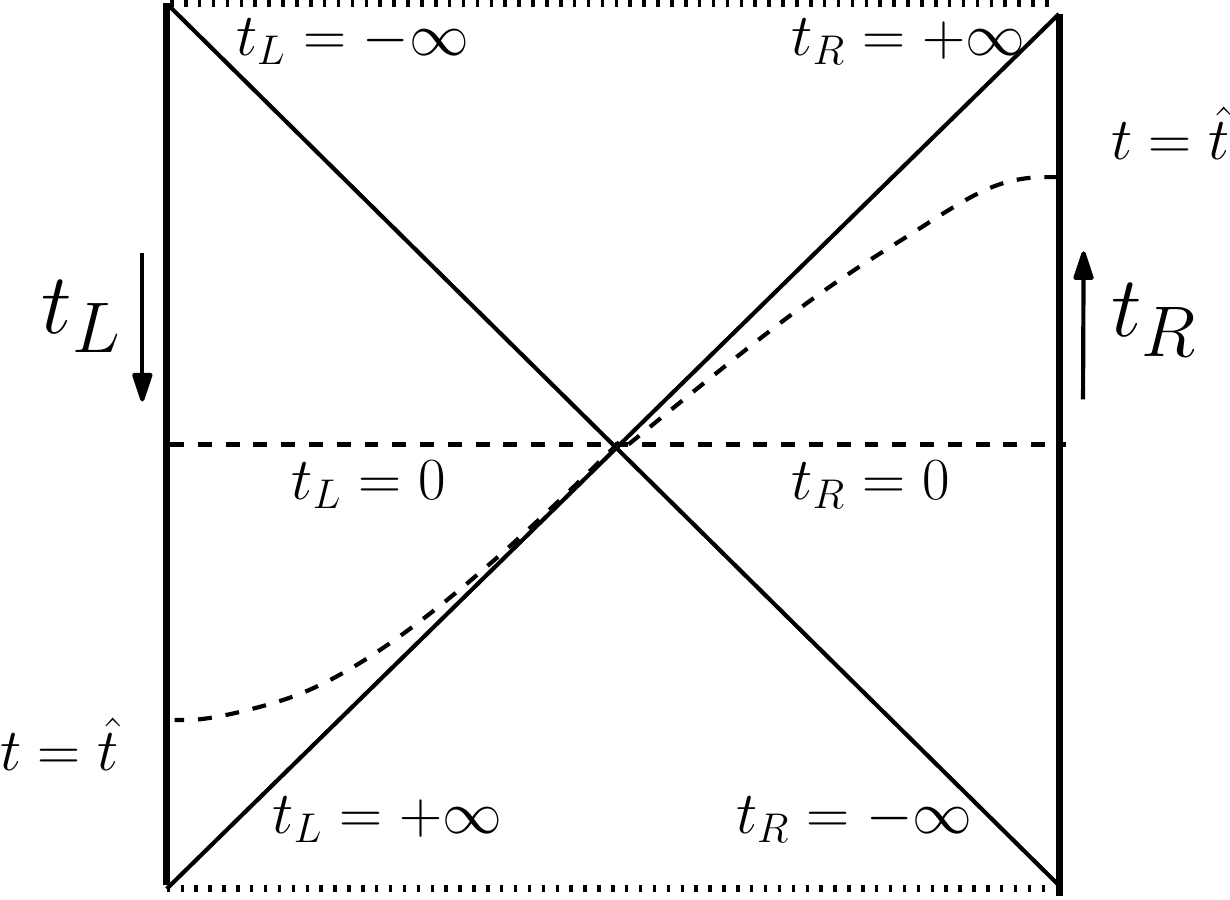} }}%
    \qquad\qquad \qquad 
    \subfloat[]{{\includegraphics[width=4.5cm]{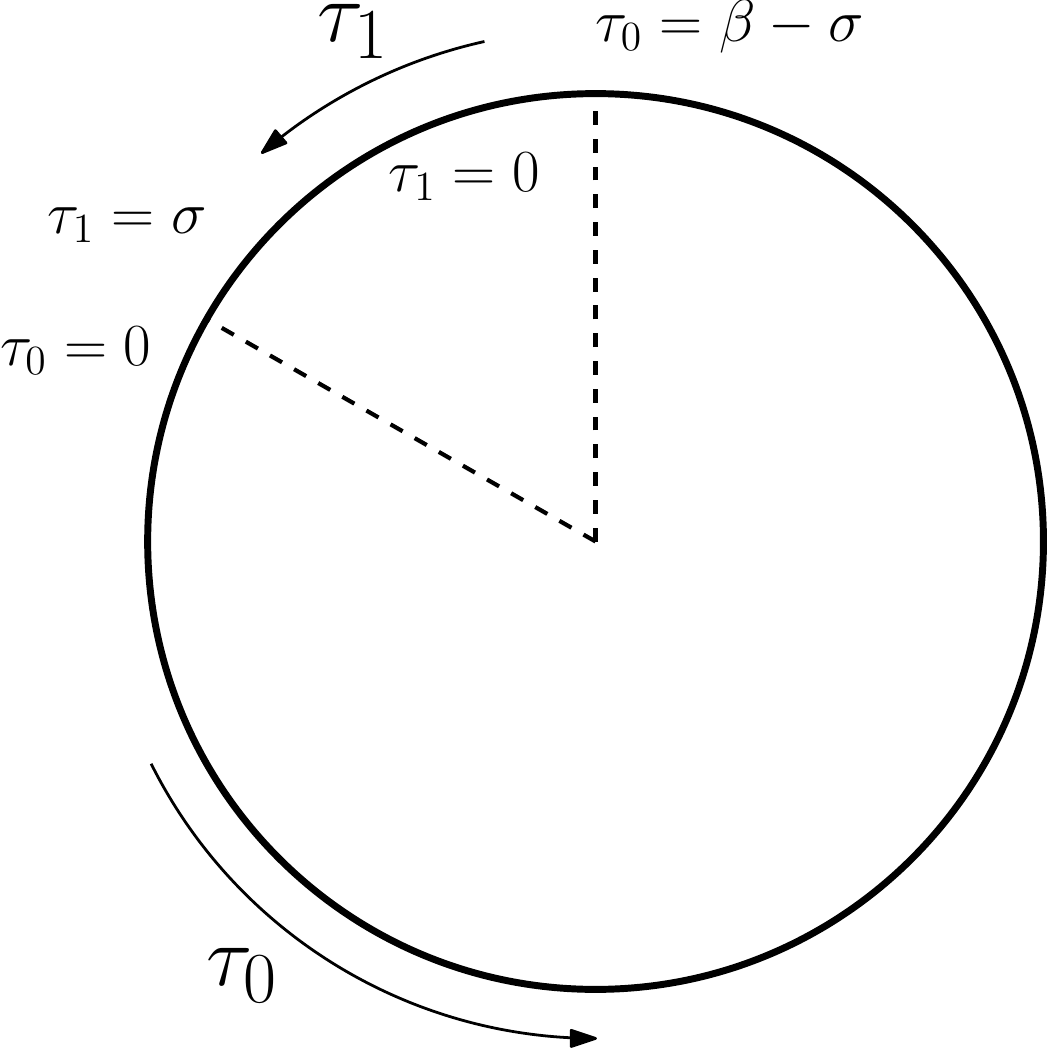} }}%
    \caption{On the left-hand side (a): an eternal AdS black hole where the arrows indicate the flow of time. We cut along the     an initial time slice at $t=0$ and  a late time slice at $t=\hat{t}$ (dashed lines), and we keep the  region in between them. On the right-hand side (b): a Euclidean  black hole in AdS with period $\beta$. We cut along finite time slices at $\tau=0$ and $\tau = \beta-\sigma$ (dashed lines).  For simplicity we have depicted the spacetimes in $2+1$ dimensions in the global time and radial coordinate.}%
    \label{fig.spacetime}%
\end{figure}

\begin{figure}[t]
\centering
 \includegraphics[width=5.7cm]{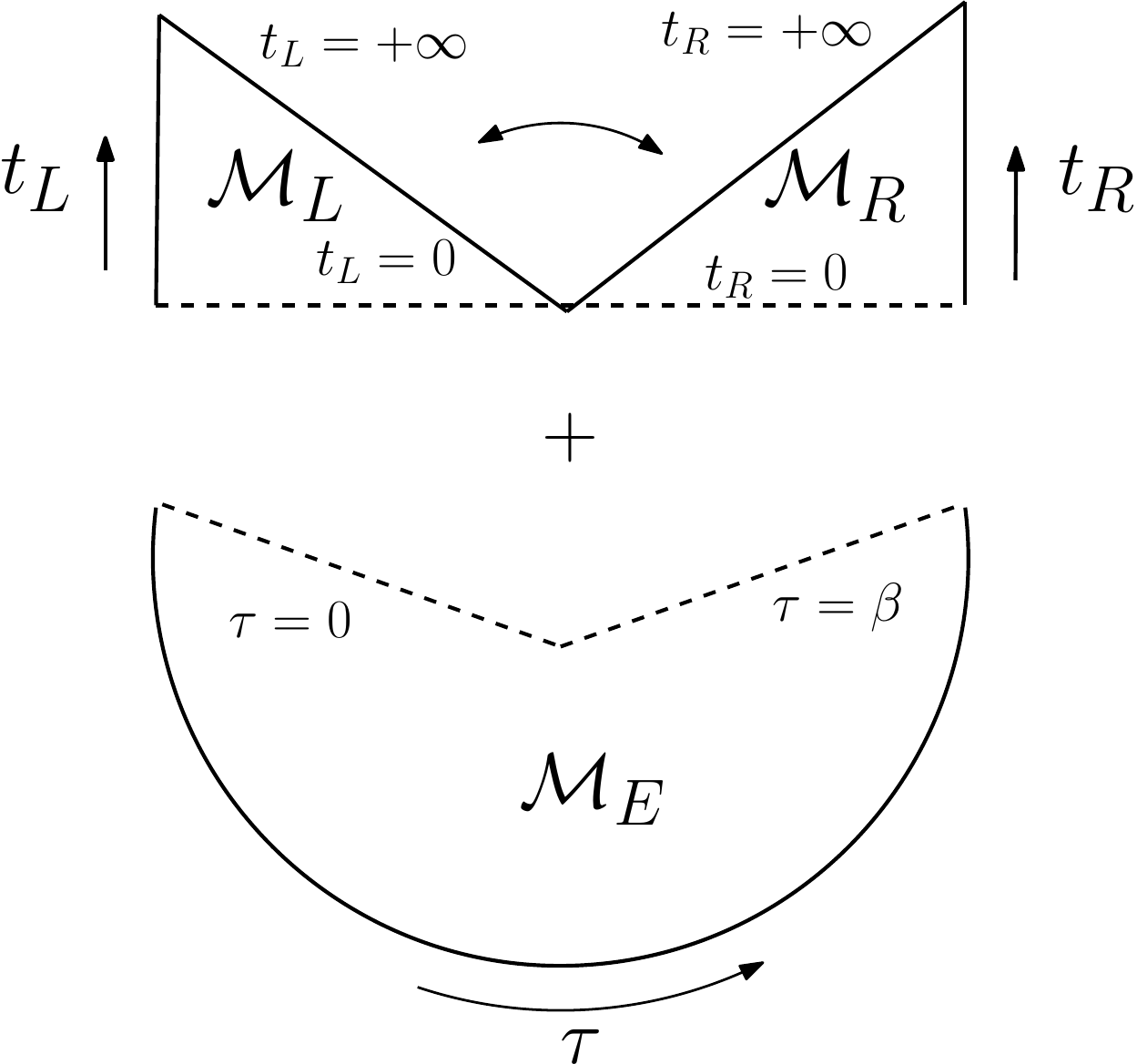}
\caption{The mixed signature bulk spacetime containing two Lorentzian regions, ${\cal M}_R$ and ${\cal M}_L$, with future horizons identified and one Euclidean black hole ${\cal M}_E$ glued smoothly together along finite time slices. Notice that in order to simplify the drawing we flipped the left segment with respect to the way it appears in fig.~\ref{fig.spacetime}, so that the left time increases now upwards.}
\label{fig.SKfinal}
\end{figure}

The bulk gravity solution dual to the field theory Schwinger-Keldysh contour of Fig.~\ref{fig.closed} for a thermal state can be  realized as follows.  The real-time field theory segments are given by the boundaries of a Lorentzian eternal black hole in AdS cut  along some finite time slices\footnote{We could as well have  chosen to place the initial state at some time in the infinite past  $t=-\infty$. This would correspond in the field theory side to the contour in  Fig.~\ref{fig.closed} extended all the way to $t=-\infty$.
} between, say,   $t=0$ and $t=\hat{t}$. 
The two imaginary time segments in Fig.~\ref{fig.closed} are given by the boundary of the analytically-continued\footnote{For simple spacetimes, as the one considered in this work, the analytic continuation to Euclidean time gives real spacetime metrics. The prescription necessitates, however, complex spacetimes when considering, say, rotating black holes, see ref. \cite{Skenderis:2008dg}. Recent works~\cite{Feldbrugge:2017kzv,Feldbrugge:2017fcc}, building on~\cite{Vilenkin:1982de,Vilenkin:1983xq,Hartle:1983ai} revived interests in constructing complex spacetimes also outside the realm of holography. Of course, the challenge is to find fully non-linear solutions of Einstein equations with matter without too restrictive symmetry assumptions on the boundary sources. One simple way to circumnavigate this demanding problem is by constructing the bulk by gluing pieces of time-translationally-invariant geometries and only in this analytically-given background explicitly solve for the actual dynamics of probe fields (gauge fields, metric perturbations, etc).\label{footnote:complexmetrics}} Euclidean AdS black hole with period~$\beta$. Cutting along two finite time slices,  at $\tau=0$ and $\tau =\beta-\sigma$ as depicted in  Fig.~\ref{fig.spacetime}, 
  we glue
 $t_L=0$ to $\tau_0=0$ and  $\tau_0=\beta-\sigma$ to $t_R=0$. Then we glue  $t_R=\hat{t}$ to 
$\tau_1=0$ and $t_L=\hat{t}$ to $\tau_1=\sigma$. By doing so, we are ensuring that the  total periodicity of the Euclidean time is $\beta$. The Schwinger-Keldysh contour of Fig.~\ref{fig.closed} is then realized  by sending $\hat{t}\rightarrow +\infty$ such that  the late time slice corresponds to  the future horizons. Moreover, we can also  choose to set  $\sigma=0$  so that the future horizons in the right and left region of the eternal black hole  are effectively identified with one another. In this way, the bulk spacetime precisely realizes the contour in Fig.~\ref{fig.closed} where the two real time segments are taken to be lying on top of each other. 
The resulting mixed signature bulk spacetime is summarized in Fig.~\ref{fig.SKfinal}.  
Two Lorentzian geometries, ${\cal M}_R$ and ${\cal M}_L$,  are glued together along  $t=+\infty$ and to a Euclidean cap ${\cal M}_0$ along $t=0$. The gluing is smooth by construction since the metric and the extrinsic curvature are identified across these finite time slices. 
A bulk spacetime of this sort appeared before in ref.~\cite{vanRees:2009rw}. What is interesting is that the black hole interior is not included in this geometrization of the thermal Schwinger-Keldysh profile in the probe approximation and will play no role in the calculation we are about to perform.

A different choice of $\sigma$ would lead to another parameterization of the Schwinger-Keldysh path integral and the resulting correlators. For example, $\sigma=\beta/2$ would correspond to the thermofield double partition function 
\be\label{eq:TFD}
Z[A_1,A_2]=\text{Tr}\left({\cal U}[A_1]\rho^{\beta/2}{\cal U}[A_2]\rho^{\beta/2}\right)
\ee
where $\rho^{\beta/2}=e^{-\frac{1}{2}\beta H}$. A parameterization of this type was effectively considered in ref.~ \cite{Herzog:2002pc} and more recently in ref.~\cite{Botta-Cantcheff:2018brv} precisely to build the holographic dual of eq.~\eqref{eq:TFD} through the real-time holography techniques of refs.~\cite{Skenderis:2008dh,Skenderis:2008dg}. 

\subsection{The degrees of freedom} 

Now that we have build the mixed signature bulk spacetime necessary to derive the holographic dual of the Schwinger-Keldysh effective action, we need to identify the bulk dual of the  relevant  low-energy degrees of freedom responsible for hydrodynamic phenomena. 

In the simple case of charge diffusion in a fixed thermal background, the field theory degrees of freedom are two scalar fields $\phi_R$ and $\phi_L$ as previously discussed around eq.~\eqref{eq:B}. 
The holographic dual of one of these quantities was first introduced in ref.~\cite{Nickel:2010pr} as a Wilson line of a probe bulk gauge field $A_M$ and subsequently covariantly generalized in ref.~\cite{deBoer:2015ija} to correspond to spacelike geodesics extending between two boundaries of the bulk spacetime
\be\label{Eq:phi}
\phi= \int_{\gamma(\lambda_1)}^{\gamma(\lambda_2)}A_{M}V^M d\lambda\,.
\ee
Here $V^{M}=dx^{M}/d\lambda$ is the tangent vector to spacelike geodesics $\gamma$ with affine parameter $\lambda$ extending between two timelike boundaries  at $(x^{\mu},\gamma(\lambda_1))$ and $(x^{\mu},\gamma(\lambda_2))$, while $M$ labels coordinates in the bulk. In particular,  the boundaries are taken to be the conformal boundary at infinity (which can be interpreted in light of the discussion in sec. \ref{S:SK} as one of the target spaces) and an intermediate timelike slice at finite radius, very close to the horizon of a black brane in AdS (interpreted as the worldvolume).

Here we are going to consider two of these Wilson lines precisely  to retain the doubling structure of the fields and sources of the Schwinger-Keldysh effective action. Starting from the interior of the spacetime close to the bifurcation horizon we extend a Wilson line towards each of the left and right conformal boundaries of the Lorentzian regions ${\cal M}_L$ and ${\cal M}_R$ in Fig.~\ref{fig.SKfinal}. These Wilson lines will correspond to the low-energy degrees of freedom $\phi_R$ and $\phi_L$ and in fact we will use the same symbols when dealing with these objects.

Let us elaborate on why eq.~\eqref{Eq:phi} is a valid choice for the low-energy degree of freedom describing charge diffusion. Consider bulk gauge transformations $\delta_{\Lambda} A_M = \partial_M\Lambda$ with parameter~$\Lambda$. The transformations which are equal at the two boundaries with  $\Lambda(\lambda_1)=\Lambda(\lambda_2)$ leave $\phi$ in \eqref{Eq:phi} invariant\footnote{Given the freedom in redefining the gauge parameter by a constant, we can equivalently set the values of these gauge transformations at the boundary to vanish.}. Under a generic bulk gauge transformations with $\Lambda(\lambda_1)\neq\Lambda(\lambda_2)$, the field $\phi$ transforms instead as a bifundamental field with respect to what can be interpreted as independent gauge transformations on the two boundaries
\be
\phi\rightarrow \phi +\Lambda(\lambda_2)-\Lambda(\lambda_1)\,,
\ee
similarly to what happens in eq.~\eqref{SK:bi}. This property is almost all we want. We still need  the field $\phi$ to transform under a restricted version of the worldvolume gauge transformations  as discussed around eq.~\eqref{cond5}. This can be achieved when one of the boundaries is pushed towards the horizon. Here, the time component of the gauge field needs to be set to zero given that otherwise the gauge connection on the corresponding Euclidean cap would be singular, see, e.g., ref.~\cite{Hartnoll:2011fn}. Thus, to avoid  the time component of the gauge field $A_t$ to acquire any non-zero value at the horizon,  only time-independent gauge transformations are allowed there\footnote{More precisely, the covariant requirement is $\oint A =0$ at the horizon. A pure gauge boundary condition $A_Mdx^M=dx^M\partial_Mf$  at the horizon with $f$ a generic function would therefore satisfy the above constraint. In this way, any gauge transformation is allowed at the horizon  as it would preserve the covariant condition above.}, precisely as in eq.~\eqref{cond5}.

 These properties make it manifest that the aforementioned Wilson lines  are continuously connected to the Goldstone modes of spontaneously broken global symmetries, see the original ref.~\cite{Nickel:2010pr} and subsequent developments in refs.~\cite{Faulkner:2010jy,deBoer:2015ija,Crossley:2015tka}. The main idea in these works is the following. It is well known that in holography global symmetry transformations of a dual field theory are represented by gauge transformations of the bulk fields that act in a desired way (i.e. as a desired global transformation) at the boundary. However, if the bulk contains two boundaries (here, as in refs.~\cite{Nickel:2010pr,Faulkner:2010jy,deBoer:2015ija,Crossley:2015tka}, the UV boundary and a surface right outside the horizon), then bulk gauge transformations induce two global symmetries -- one for each boundary. However, the connectedness of the bulk makes independent global transformations impossible without preserving the solution, which is a clear manifestation of a spontaneous symmetry breaking. In the present case, the relevant symmetry breaking pattern is $U(1)\times U(1)$ into the diagonal~$U(1)$, which gives rise to a single Goldstone boson~\eqref{Eq:phi}.

To make sure that the quantity in \eqref{Eq:phi} is well defined, we require  
$\phi$ to be invariant under infinitesimal variations of the bulk gauge field $\delta A_M$. The latter  can be separated into longitudinal and perpendicular  contributions to the tangent vector $V^M$,  
\be
\delta A_M = V_MV^N\delta A_N+ P_{M}^N \delta A_N \,,
\ee
where $P_{MN}=g_{MN}-V_MV_N/V^2$ is the projector onto the orthogonal directions, satisfying $P_{MN}V^N=0$. It is immediate to see that orthogonal variations do not modify $\phi$ in eq.~\eqref{Eq:phi} while longitudinal ones  do, unless they are of a pure gauge form $V^M\delta A_{M}=V^M\partial_M h$ with $h$ a generic function. In what follows we will thus work with a constrained variational principle where longitudinal variations of the bulk gauge field are set to zero 
\be
V^M\delta A_M =0\,.
\ee
Given the action for a probe gauge field
\be
\label{eq:action}
S=-\frac{1}{4g_A^2}\int d^{d+1}x\sqrt{-g}F_{MN}F^{MN}\,,
\ee
where $g_{A}$ is the effective gauge coupling and $F_{MN}$ is the bulk electromagnetic field strength, the resulting equations of motion can also be separated into longitudinal and perpendicular contributions, 
\be
\delta S= \frac{\delta S}{\delta A_M}\delta A_M \sim \left(V_M\nabla_NF^{NM}\right)V^Q\delta A_Q+\left(P^Q_M\nabla_NF^{NM}\right)\delta A_Q \,.
\ee
In practice, to implement the constrained variational principle and to have a well defined dynamical field $\phi$ in eq.~\eqref{Eq:phi}, we shall not solve the equation of motion along the tangent vector $V_M\nabla_{N}F^{NM}=0$ which is dubbed as the {\it constraint equation}.

A similar analysis to the one considered in this section goes through for the case of linear fluid dynamic equations, thus, holographically, for bulk metric perturbations, see, e.g., refs.~\cite{deBoer:2015ija,Crossley:2015tka} for a definition of the corresponding low-energy fluid degrees of freedom in holography. This setup, however, lies outside of the scope of the present work.


\section{The effective action for diffusion from holography}
\label{S:diff}


In this section we are going to consider in more detail the example of a  conserved charge current in (1+3)-dimensional strongly coupled conformal field theories at large $N$. Through holography, this system corresponds on the gravity side to the low-energy dynamics of a probe $U(1)$ gauge field in a fixed spacetime at finite temperature given by an AdS$_5$ black brane metric.

In Lorentzian signature the background metric can be written using Schwarzschild-like coordinates 
\ba\label{E:lor}
ds^2&=&\frac{(\pi\, T L)^2}{u}\left(-f(u)\,dt^2+dx^2+dy^2+dz^2\right)+\frac{L^2du^2}{4\,u^2f(u)},
\ea
where
\ba
f(u)=1-u^2
\ea
is the emblackening factor, $T$ is the Hawking temperature and $L$ is the AdS curvature radius. The radial coordinate $u$ stretches between the bifurcation surface at $u=1$ and the conformal boundary located at $u=0$. The future horizon corresponds to $u = 1$ and $t \rightarrow +\infty$, see, e.g., ref.~\cite{Bhattacharyya:2008mz}. The Euclidean black brane  metric can be obtained from eq.~\eqref{E:lor} by a Wick rotation of the time coordinate $t=- i\,\tau$,
\ba\label{E:eucld}
ds^2&=&\frac{(\pi \,T L)^2}{u}\left(f(u)\,d\tau^2+dx^2+dy^2+dz^2\right)+\frac{L^2du^2}{4\,u^2f(u)}\,.
\ea
As described in the previous section we will make  use of  both  of these metrics. Two Lorentzian spacetimes of the form \eqref{E:lor}, corresponding to the  left  and right  parts of the eternal black hole ${\cal{M}}_L$ and ${\cal{M}}_R$, are glued to the Euclidean spacetime  \eqref{E:eucld} parameterizing the Euclidean cap  ${\cal{M}}_E$, see Fig.~\ref{fig.SKfinal}. 
To do so, one needs to ensure that the continuity conditions are satisfied across the gluing surfaces. This is easily achieved in our fixed background since the gluings are performed  along fixed time slices, where the induced metric and the extrinsic curvature of the metrics \eqref{E:eucld} and \eqref{E:lor} coincide. 
Notice that in building the mixed signature background there are three sets of coordinates which we will parameterize with subscripts $R$, $L$ and $E$ when referring to ${\cal M}_R$,  ${\cal M}_L$ and ${\cal M}_E$  respectively. However, to avoid cluttering of notation, we will also often omit these subscripts  when their meaning is clear from the context.

The action of the probe $U(1)$ gauge field $A_M$ in Lorentzian signature is given by eq.~\eqref{eq:action} with $d = 1+3$. It is convenient to work in Fourier space where
\be
\label{eq.AMfourier}
A_{M}(t,x,u)=\int \frac{d\omega \,d\vec{k}}{(2\pi)^4}\,e^{-i\,\omega\, t+i\,k\,x}A_{M}(u)\,,
\ee
and, without loss of generality, we have chosen to align  the wave vector along the $x$ direction $\vec{k}=(k,0,0)$. In the following we will suppress, as was the case in eq.~\eqref{eq.AMfourier}, the dependence of functions on frequency and momentum, when it follows from the context. The components of the gauge vector $A_{M}$ can be organized according to their transformation properties with respect to the residual symmetry group $O(2)$  of rotations in the plane transverse to momentum as follows
\begin{subequations}
\begin{align}
\text{longitudinal channel:} &\qquad A_t,\quad A_x,\quad A_u \label{eq.longchann}\,,\\
\text{transverse channel:}&\qquad A_{\alpha}\qquad\text{with}\qquad \alpha=y,z\,\label{eq.transvchann}.
\end{align}
\end{subequations}
As a result, the equations of motion, 
\be\label{eq:EOM}
\nabla_{M} F^{MN}=0,
\ee
separate between the two channels. In the reminder of this work we will consider  the longitudinal channel~\eqref{eq.longchann} as it is the only one which exhibits a non-trivial hydrodynamic behavior. 

We find it convenient to re-parameterize the fields as follows
 \be\label{eq:redB}
 A_u(u)=-\phi'(u)\,, \qquad B_t(u)=A_t(u)-i\,\omega\,\phi(u)\,,\qquad B_{x}(u)=A_x(u)+i \,k\, \phi(u)\,,
 \ee
 where $'$ denotes the derivative with respect to the radial coordinate $u$.
  The  field $\phi$ is a non-local Wilson line
  \be\label{eq:Wilson}
  \phi(u)=-\int^{u}_{\delta} A_{u}(\tilde{u}) \,d\tilde{u}
  \ee 
  extending along geodesics at fixed $(t,\vec{x})$ between two  hypersurfaces located at fixed radii $\delta$ and  $u$. When these boundaries are stretched between the horizon  $\delta=1$ and the conformal boundary  $u=0$, as it has been anticipated in the previous section around eq.~\eqref{Eq:phi}, the field~\eqref{eq:Wilson} can be interpreted as the  holographic dual of the low-energy degree of freedom capturing the  long-lived excitations of charged matter of the dual field theory side, see ref.~\cite{Nickel:2010pr} and subsequent refs.~\cite{deBoer:2015ija,Crossley:2015tka}. The only difference between eq.~\eqref{eq:Wilson} and the field appearing in the definition~\eqref{eq:B} is  that the former is defined in Schwarzschild coordinates, thus boundary (target space) coordinates, while the latter is defined on the worldvolume. We could work with combinations in \eqref{eq:redB} pulled back to the horizon (i.e.~the~worldvolume) but we choose not to. Such combinations would differ only by an overall multiplicative factor which will be cancelled anyway  given that we are interested in the physical quantities defined  on the target spaces via pushforwards, as in eq.~\eqref{eq:push}.
   
Under bulk $U(1)$ gauge transformations $A_M\rightarrow A_M+\partial_M \Lambda$,  the field $\phi$ defined in eq.~\eqref{eq:Wilson} transforms as a bifundamental field under what can be interpreted as independent gauge transformations on the two fixed radius hypersurfaces
    \be\label{eq:bif}
  \phi(u)\rightarrow \phi(u)-\Lambda(u)+\Lambda(\delta)\,,
  \ee
  similarly to eq.~\eqref{SK:bi}.
  The Wilson line \eqref{eq:Wilson} is clearly  invariant under   diagonal combinations of the boundary gauge transformations, those for which  $\Lambda(u)=\Lambda(\delta)$, and transforms non trivially under the remaining transformations for which $\Lambda(u)-\Lambda(\delta)\neq 0$. 
As discussed shortly before, it is  this reason why the field in eq.~\eqref{eq:Wilson} is sometimes referred to as the Goldstone mode of the broken $U(1)\times U(1)$ global symmetry subgroup of the gauge invariances of the two boundaries down to the diagonal combination thereof.

Finally, the equations of motion \eqref{eq:EOM} for the newly redefined variables \eqref{eq:redB} in the Lorentzian background \eqref{E:lor}  are given by
 \begin{subequations}
\begin{align}
 &B_t''-\frac{\tilde{k}^2}{u\,f}B_t-\frac{\tilde{k}\,\tilde{\omega}}{u\,f}B_x=0\,,\label{eq:Bt}\\
 &B_x''+\frac{f'}{f}B_x'+\frac{\tilde{\omega}^2}{u\,f^2}B_x+\frac{\tilde{k}\,\tilde{\omega}}{u\,f^2}B_t=0\,,\label{eq:Bx}\\
 &\tilde{\omega}\, B_t'+\tilde{k}\,f\,B_x'=0\,,\label{E:constr}
\end{align}
\end{subequations}
 where  the dependence on $u$ is implied and we have introduced the   rescaled frequency and momentum 
 \be
 \tilde{\omega}=\frac{\omega}{2\pi T}\qquad \text{and}\qquad \tilde{k}=\frac{k}{2\pi T}\,.
 \ee
 The first two equations \eqref{eq:Bt} and \eqref{eq:Bx} are the dynamical equations for $B_t$ and $B_x$ while the last equation \eqref{E:constr} is a constraint, i.e. the field equation projected along the geodesic on which the Wilson line \eqref{eq:Wilson} is defined.  Our primary goal is to solve the dynamical equations while leaving the constraint unsolved. This will allow us to compute a partially on-shell bulk action which will be interpreted as the local effective action for the low-energy dynamics of charged matter in the dual field theory.


\subsection{The bulk piecewise solution}
\label{S:IRaction}


Let us now derive the bulk solution by solving the field equations in the various parts of the spacetime and patching them together along the gluing surfaces at constant time. 
 Performing the near horizon limit first $u\rightarrow 1$ and then the hydrodynamic limit
\be
\tilde{\omega}, \, \tilde{k}\ll 1\,,
\ee
the equations \eqref{eq:Bt} and \eqref{eq:Bx}  admit a simple solution  of the form 
\be\label{eq:nearhor}
B_t=a_c+a_l(1-u)\qquad \text{and} \qquad B_x=a_+(1-u)^{+i\tilde{\omega}/2}+a_-(1-u)^{-i\tilde{\omega}/2}\,,
\ee
where $a_c$, $a_l$ and $a_{\pm}$ are integration constants. The time component of $B_t$ has a constant and linear behavior at the horizon, while $B_x$ is a superposition of an ingoing and outgoing mode.

At higher orders in derivatives, the equations \eqref{eq:Bt} and \eqref{eq:Bx} are coupled and it is plausible  to work with the following ansatz
\ba\label{Eq:ansatzB}
B_{t}=a_{c}\,m+a_{l}\,n+a_+g_++a_-g_-\qquad \text{and} \qquad
B_{x}=a_c\,p+a_l\,q+a_{+}h_++a_{-}\,h_-\,,\quad
\ea
where $m=m(\tilde{\omega},\tilde{k},u)$, $n=n(\tilde{\omega},\tilde{k},u),\dots$ are  linearly dependent mode solutions. Only 4 solutions are independent, in fact by inserting  expressions \eqref{Eq:ansatzB} into eq.~\eqref{eq:Bt} we can immediately show that one can express $p$, $q$, $h_{\pm}$ in terms of $m$, $n$, $g_{\pm}$. We require  that near the  horizon the solution \eqref{Eq:ansatzB} reproduces the behavior seen in eq.~\eqref{eq:nearhor}, thus 
 \begin{subequations}
\begin{align}
 \label{eq:modes}
& m=1+\lambda\, m^{(1)}+\dots\,,\\
& n=(1-u)\left(1+\lambda\, n^{(1)}+\dots\right)\,,\\
&g_{\pm}=\lambda\, g_{\pm}^{(1)}+\dots\,, \\
&p=\lambda\, p^{(1)}+\dots\,,\\ 
&q=(1-u)\left(\lambda\, q^{(1)}+\dots\right)\,,\\
&h_{\pm}=(1-u)^{\pm\tilde{\omega}/2}\left(1+\lambda\, h_{\pm}^{(1)}+\dots\right)\,,
\end{align} 
\end{subequations}
where $\lambda$ is a bookkeeping parameter counting the number of derivatives and dots denote higher order terms in the hydrodynamic expansion.
Thus, the general bulk solution in the various parts of the spacetime can be parameterized by 12 independent  coefficients, 8 in the Lorentzian parts given by $a^{R/L}_c$, $a_l^{R/L}$ and $a_{\pm}^{R/L}$,
\begin{align}
 \begin{split}
& B_{s\,t}(t_s,x,u)=\int \frac{d\omega\, d\vec{k}}{(2\pi)^4}e^{-i\,\omega\, t_s+i\,k\,x}\left(a^s_{c}\,m+a^s_{l}\,n+a^s_+g_++a^s_-g_-\right)\,,\\
& B_{s\,x}(t_s,x,u)=\int \frac{d\omega\, d\vec{k}}{(2\pi)^4}e^{-i\,\omega\, t_s+i\,k\,x}\left(a^s_c\,p+a^s_l\,q+a^s_{+}h_++a^s_{-}\,h_-\right)\,,
 \end{split}
 \end{align}
 where $s=R,\,L$ and 4 more in the Euclidean manifold given by $a_c^E$, $a_l^E$ and $a_{\pm}^E$,
\begin{align}
 \begin{split}
& B_{E\,\tau}(\tau,x,u)=\int \frac{d\omega\, d\vec{k}}{(2\pi)^4}e^{-\omega\, \tau+i\,k\,x}\left(a^E_{c}\,m+a^E_{l}\,n+a^E_+g_++a^E_-g_-\right)\,,\\
& B_{E\,x}(\tau,x,u)=\int \frac{d\omega\, d\vec{k}}{(2\pi)^4}e^{-\omega\, \tau +i\,k\,x}\left(a^E_{c}\,p+a^E_{l}\,q+a^E_+h_++a^E_-h_-\right)\,.
 \end{split}
 \end{align}
 
We now impose Dirichlet boundary conditions on the conformal boundaries of the two Lorentzian parts of the spacetime and trivial (i.e. vanishing) boundary conditions on the Euclidean cap,
\begin{align}
\begin{split}\label{eq:Dirichlet}
& B_{s\,t}(t_s,x,u=0)=B_{s\,t}(t_s,x)\,,\qquad B_{s\,x}(t_s,x,u=0)=B_{s\,x}(t_s,x)\,,\\
& B_{E\,\tau}(\tau,x,u=0)=0\,,\qquad B_{E\,x}(\tau,x,u=0)=0\,
\end{split}
\end{align}
where, again, $s = R,\,L$. We refrained from using another symbol for the sources, the distinction between the bulk gauge field and its boundary value should be clear from the context. 
Non-vanishing non-normalizable modes in the Euclidean part of the spacetime can be interpreted as considering excited initial states, see, e.g., ref.~\cite{Botta-Cantcheff:2015sav,Christodoulou:2016nej}. Here, instead, we have restricted ourselves to the case where the initial state is simply thermal. The conditions \eqref{eq:Dirichlet} imply relations among the integration constants
\begin{align}
\begin{split}
\label{eq:conddir}
& a_c^s m^0+a_l^s n^0 +a^s_+g_{+}^0+a^s_-g^{0}=B_{s\,t}\,,\qquad  a_c^s p^0+a_l^s q^0 +a^s_+h_{+}^0+a^s_-h^{0}=B_{s\,x}\,,\\
& a_c^E m^0+a_l^E n^0 +a^E_+g_{+}^0+a^E_-g^{0}=0\,,\qquad 
 a_c^E p^0+a_l^E q^0 +a^E_+h_{+}^0+a^E_-h^{0}=0\,,
\end{split} 
\end{align}
and we have introduced the shortcut notation where, for example, $n^0 = n(\tilde{\omega},\tilde{k},u=0)$.
Using these expressions it is possible to fix 6 of the integrations constants, say $a_l^R$, $a_l^L$, $a_l^E$, $a_-^R$, $a_-^L$ and $a_-^E$, in terms of the other ones.

Another condition on the coefficients arises requiring the Euclidean cap not to have a singular gauge connection at the horizon, see, e.g., ref.~\cite{Hartnoll:2011fn}. This can be achieved by demanding that the time component of the gauge field vanishes at the horizon as we have discussed in the previous section.
 We thus impose
\be\label{eq:conconst}
a_c^E=0\,.
\ee
Continuity conditions across the gluing surfaces imply that the same constraints should hold in the Lorentzian parts of the bulk spacetime given that these modes all meet at the bifurcation point,
\be\label{eq:conconst2}
a_c^R=0\,,\qquad a_c^L=0\,.
\ee

The next step in constructing the full solution in our spacetime depicted in Fig.~\ref{fig.SKfinal} is to impose continuity of the gauge field along the surfaces at which the Lorentzian parts of the spacetime are glued to the Euclidean black hole.
These conditions are simply given by 
\begin{align}
\begin{split}
\label{eq:Btmatch}
&B_{L\,t}(t_L=0,x,u)=B_{E\,\tau}(\tau=0,x,u)\,,\qquad B_{R\,t}(t_R=0,x,u)=B_{E\,\tau}(\tau=\beta,x,u)\,,\\
&B_{L\,x}(t_L=0,x,u)=B_{E\,x}(\tau=0,x,u)\,,\qquad B_{R\,x}(t_R=0,x,u)=B_{E\,x}(\tau=\beta,x,u).
\end{split} 
\end{align}
and imply the relations 
\begin{align}
\begin{split}
\label{eq:condtime}
a_+^E = a_+^L\,,\qquad a_+^R = e^{-\omega/T}a_+^L\,,
\end{split} 
\end{align}
where $\beta=1/T$ is the period of the Euclidean cap. 
The requirement that also the time derivative of the fields across the gluing surfaces is continuous does not impose additional constraints.

To derive eq.~\eqref{eq:condtime} we used the fact that  source insertions $B_{s\,t}$ and $B_{s\,x}$  are vanishing at $t_L=0$ and $t_R=0$  to match the boundary condition on the  Euclidean cap and have support only at later times.  After imposing the relations \eqref{eq:conddir}, \eqref{eq:conconst},  \eqref{eq:conconst2} and \eqref{eq:condtime},  the combinations in \eqref{eq:Btmatch} depend only on the external sources, as they should. For~example,~we~have
\begin{align}
\begin{split}
\label{eq:convergence}
 \hspace{-7 pt}&B_{R\,x}(t_L=0,x,u)-B_{E\,x}(\tau=1/T,x,u)=\int \frac{d\omega\,d\vec{k}}{(2\pi)^4}e^{ikx}\left(\Delta_{-}^{xt}B_{R\,t}(\omega,k)+\Delta_-^{xx} B_{R\,x}(\omega,k)\right) = \\
 \hspace{-7 pt}&\int dt^{\prime}\, dx^{\prime}\left( B_{R\,t}(t^\prime,x^\prime)\int \frac{d\omega\,d\vec{k}}{(2\pi)^4}e^{+i\omega t^{\prime}}e^{-ik(x^\prime - x)}\Delta_-^{xt}+B_{R\,x}(t^\prime,x^\prime)\int \frac{d\omega\,d\vec{k}}{(2\pi)^4}e^{+i\omega t^{\prime}}e^{-ik(x^\prime - x)}\Delta_-^{xx}\right)\\
\end{split}
 \end{align}
 with $\Delta_{-}^{xt}$ and $\Delta_-^{xx}$ being functions of the modes $m, h_+ \dots$ which we will define shortly.
The combination in eq.~\eqref{eq:convergence} is vanishing for  $t^{\prime}>0$  as long as $\Delta_{-}^{xt}$ and $\Delta_-^{xx}$ have no poles in the upper half of the complex $\omega$-plane where the integral is convergent. We have verified up to second order in the derivative expansion that $\Delta_{-}^{xt}$ and $\Delta_-^{xx}$ indeed satisfy the above requirement. It is also natural to expect that this feature holds also at higher orders in derivatives given that the propagators $\Delta_-$ have essentially an ingoing behavior near the horizon, although we have not verified this statement explicitly.

Finally, let us  impose a gluing condition  on the late time surface  $\hat{t}$ which is taken to $\hat{t}\rightarrow +\infty$ in such a way that it coincides with the future horizons. The precise position of this late time hypersurface is not so important as long as it is later than any source insertion. Thus, we require
\be\label{eq:Bmatch2}
B_{L\,t}(t_L=\hat{t},x,u)=B_{R\,t}(t_R=\hat{t},x,u)\,,\qquad B_{L\,x}(t_L=\hat{t},x,u)=B_{R\,x}(t_R=\hat{t},x,u)\,,
\ee
which   give
\be\label{eq:apL}
a_{+}^L=(1+\bar{n}(\omega))\frac{(B_{L\,x}-B_{R\,x})n^0-(B_{L\,t}-B_{R\,t})q^0}{h_+^0n^0-g_+^0 q^0}\,,
\ee
where we have defined the Boltzmann distribution
\be
\bar{n}(\omega)=\frac{1}{e^{\omega/T}-1}\,.
\ee
In obtaining eq.~\eqref{eq:apL} we used similar arguments as the ones appearing around eq.~\eqref{eq:convergence}. The late time surface is located at times 
 $\hat{t}$  bigger than any possible other time where the sources might have been inserted making  the domain of convergence of the integrals involved in eq.~\eqref{eq:Bmatch2} being the lower half of the  complex  $\omega$-plane. We will have to make sure that after imposing the condition on the coefficients \eqref{eq:apL} the argument of the integrals has poles at most in the upper half complex $\omega$-plane. 

Having fixed all the integration constants in terms of the boundary sources, we have succeeded in finding a regular formal bulk  solution interpolating between the Lorentzian and Euclidean manifolds as a function of  the sources $B_{L\,t}$, $B_{R\,t}$, $B_{L\, x}$ and $B_{R\, x}$ inserted on the left and right conformal boundaries. To summarize, the solution can be parameterized as follows 
\begin{align}
\begin{split}
\label{sol}
B_{s\,\mu}(t_s,x,u)=&\int \frac{d\omega\, d\vec{k}}{(2\pi)^4}\, e^{-i\omega t_s+ikx}\sum_{m=R,L}\sum_{\nu=t,x}\Delta^{\mu\nu}_{sm}(\tilde{\omega},\tilde{k},u) B_{m\, \nu}(\tilde{\omega},\tilde{k})\,,\\
B_{E\,\mu}(\tau,x,u)=&\int \frac{d\omega\, d\vec{k}}{(2\pi)^4}\, e^{-\omega\tau+ikx}\sum_{m=R,L}\sum_{\nu=t,x}\Delta^{\mu\nu}_{Em}(\tilde{\omega},\tilde{k},u)B_{m\,\nu}(\tilde{\omega},\tilde{k})\,,
\end{split}
\end{align}
where $s=R,L$, $\mu=t,x$ (or $\mu=\tau,x$) and $\Delta_{sm}^{\mu\nu}$ and  $\Delta_{Em}^{\mu\nu}$ are the bulk to boundary propagators. For example, $\Delta_{LL}^{tt}$ is the propagator which determines the bulk gauge field $B_{L\,t}$ in ${\cal M}_L$ under the influence of the source $B_{L\,t}$ inserted at the conformal boundary of the same manifold. On the other hand, $\Delta_{LR}^{tt}$ is the bulk to boundary propagator for the same bulk gauge field $B_{L\,t}$ in  ${\cal M}_L$, depending now on the source  $B_{R\,t}$ inserted on the conformal boundary of ${\cal M}_R$ instead.
 The bulk to boundary propagators are explicitly given by 
\begin{align}
\begin{split}\label{eq:delta1}
& \Delta_{LL}^{\mu\nu}=  -\bar{n}(\omega)\Delta_{-}^{\mu\nu}+(1+\bar{n}(\omega))\Delta_{+}^{\mu\nu}\,,\qquad 
\Delta_{LR}^{\mu\nu}=  (1+\bar{n}(\omega))\left(\Delta_{-}^{\mu\nu}-\Delta_{+}^{\mu\nu}\right)\,,\\
& \Delta_{RL}^{\mu\nu}=  -\bar{n}(\omega)\left(\Delta_{-}^{\mu\nu}-\Delta_{+}^{\mu\nu}\right)\,,\qquad
\Delta_{RR}^{\mu\nu}=  (1+\bar{n}(\omega))\Delta_{-}^{\mu\nu} -\bar{n}(\omega)\Delta_{+}^{\mu\nu}\,,  \\
& \Delta_{EL}^{\mu\nu}=-(1+\bar{n}(\omega))\left(\Delta_{-}^{ij}-\Delta_{+}^{\mu\nu}\right)\,,\qquad \Delta_{ER}^{\mu\nu}=(1+\bar{n}(\omega))\left(\Delta_{-}^{\mu\nu}-\Delta_{+}^{\mu\nu}\right)\,, 
\end{split}
\end{align}
where we have defined the combinations
\begin{align}
\begin{split}
\label{eq:deltas}
&\Delta^{tt}_{\pm}=\frac{h_{\pm}^0n-q^0g_{\pm}}{h_{\pm}^0n^0-g_{\pm}^0q^0}\,,\qquad \Delta^{tx}_{\pm}=\frac{n^0g_{\pm}-g_{\pm}^0n}{h_{\pm}^0n^0-g_{\pm}^0q^0}\,,\\
&\Delta^{xt}_{\pm}=\frac{h_{\pm}^0q-q^0h_{\pm}}{h_{\pm}^0n^0-g_{\pm}^0q^0}\,,\qquad \Delta^{xx}_{\pm}=\frac{n^0h_{\pm}-g_{\pm}^0q}{h_{\pm}^0n^0-g_{\pm}^0q^0}\,.
\end{split} 
\end{align}

Before proceeding let us first observe a simple fact. When the external sources are set to be equal to one another on the two sides of the Lorentzian spacetime $B_{L\,x}=B_{R\,x}=B_{x}$ and $B_{L\,t}=B_{R\,t}=B_{t}$, the bulk gauge  field in ${\cal M}_R$ depends only on the combination $\Delta^{\mu\nu}_-$, 
\begin{align}
 \begin{split}
 \label{eq:Deltam}
B_{R\,\mu}(t_s,x,u)\bigg|_{B_{R}=B_L=B}= &\int \frac{d\omega\, d\vec{k}}{(2\pi)^4}\, e^{-i\omega t_R+ikx}\sum_{\nu=t,x}(\Delta^{\mu\nu}_{RL}+\Delta^{\mu\nu}_{RR})B_{\nu}\\
=& \int \frac{d\omega\, d\vec{k}}{(2\pi)^4}\, e^{-i\omega t_R+ikx}\sum_{\nu=t,x}\Delta^{\mu\nu}_{-}B_{\nu}\,.
\end{split}
\end{align}
We will see in what follows that $\Delta^{\mu\nu}_{-}$ is related to ingoing modes  on the right part of the bulk spacetime ${\cal M}_R$. Thus,  as it has been previously observed in ref.~\cite{vanRees:2009rw}, selecting the sector with equal sources on the boundary is equivalent to selecting ingoing mode solutions and therefore a retarded propagation. This is not surprising and reflects the well known prescription to compute  retarded correlators in holography by imposing ingoing boundary conditions at the horizon~\cite{Son:2002sd}.

Conversely, when the sources are set to be equal to one another up to a shift by $\beta$ in imaginary time, $B_{R\,t}=e^{-\omega/T}B_{L\,t}={B}_{t}$ and $B_{R\,x}=e^{-\omega/T}B_{L\,x}={B}_{x}$, we have 
\be
 \label{eq:Deltap}
B_{R\,{\mu}}(t_R,x,u)\bigg|_{B_{R}=e^{-\omega/T}B_{L}={B}}=\int \frac{d\omega\, d\vec{k}}{(2\pi)^4}\, e^{-i\omega t_R+ikx}\sum_{\nu=t,x}\Delta^{\mu\nu}_+B_{\nu}\,,
\ee
where 
the combination $\Delta^{\mu\nu}_+$ will be  related to the outgoing mode in the bulk. Thus, this  subsector  is equivalent to selecting outgoing mode solutions. We observe, therefore, a  parallel between the two subsectors for which the Schwinger-Keldysh path integral for thermal states becomes trivial as discussed in sec.~\ref{S:SK} and the distinction between ingoing and outgoing modes in the bulk.

Let us conclude this section by commenting on a couple of important points.
Notice that  we have been cavalier in performing Fourier transforms. In fact, given that the sources have support only on a half time line between $t=0$ and $\hat{t}\rightarrow\infty$, the appropriate transformation would have been instead the Laplace transform. By working with Fourier transforms, we have abandoned the idea of finding a unique solution. We will be content here with determining $a$ solution and not to {\it uniquely} determine it. The reason is that we could have as well extended the Lorentzian manifold from  $t=0$ all the way to $t=-\infty$ and worked with a spacetime where time runs from $t=-\infty$ to $t=+\infty$. The initial state, i.e. the Euclidean cap, would have needed to be glued to the slice $t=-\infty$. We have verified that the resulting solution is equivalent to the one presented in the main text and that in this case it is unique as there are no normalizable modes compatible with the gluing conditions.

\subsection{The holographic Schwinger-Keldysh effective action}

Now that we have the formal bulk piecewise solution \eqref{sol} at our disposal, we are ready to derive the Schwinger-Keldysh effective action describing  diffusion of a $U(1)$ charge current in the dual conformal field theory in a saddle point approximation and  up to the quadratic order in the fields and sources.

We first compute the bulk action for the Maxwell field \eqref{eq:action} partially on-shell, that is using only the dynamical equations of motion given by eqs.~\eqref{eq:Bt} and \eqref{eq:Bx}. The resulting action is simply given by a term of the form
\ba\label{eq:Z:onshell}
&&\hspace{-10 pt}S_{\rm onshell}=\frac{\pi^2 T^2 L}{g_A^2}\int_{\partial {\cal M}} \frac{d\omega\, d\vec{k}}{(2\pi)^4}\,\left(B_t(-\tilde{\omega},-\tilde{k},u)\,\partial_uB_t(\tilde{\omega},\tilde{k},u)-f(u)B_x(-\tilde{\omega},-\tilde{k},u)\,\partial_uB_x(\tilde{\omega},\tilde{k},u)\right),\nn\\
&&
\ea
where $\partial {\cal M}$ is the boundary of the manifold ${\cal M}$. Secondly, we evaluate the on-shell action on the boundary of the spacetime by taking the following combination 
\be\label{eq:Sonshell}
i\,S_{eff}=i\,S_{\rm onshell}\bigg|_{u_R=0} -i\,S_{\rm onshell}\bigg|_{u_L=0} -S_{\rm onshell}\bigg|_{u_E=0}\,,
\ee
where the minus sign in front of the contribution of the left-hand side of the Lorentzian spacetime comes from the fact that we take time to run backwards in ${\cal M}_L$ to match the field theory contour in Fig.~\ref{fig.closed}. Given that there are no sources inserted at the conformal boundary of the Euclidean cap, the on-shell action corresponding to the Euclidean region will be actually vanishing.

Inserting the solution~\eqref{sol} we found in the previous Section into the expression  \eqref{eq:Sonshell} leads to
\begin{align}
\begin{split}
\label{eq:SKaction}
\hspace{-15 pt}S_{eff}=&\int \frac{d\omega\, d\vec{k}}{(2\pi)^4}\sum_{\mu=t,x}\sum_{\nu=t,x}\bigg(
B_{a\,\mu}(-\tilde{\omega},-\tilde{k})\Pi_{-}^{\mu \nu}(\tilde{\omega},\tilde{k})B_{r\,\nu}(\tilde{\omega},\tilde{k})+B_{r\,\mu}(-\tilde{\omega},-\tilde{k})\Pi_{+}^{\mu \nu}(\tilde{\omega},\tilde{k})B_{a\,\nu}(\tilde{\omega},\tilde{k})\\
&\qquad +\frac{1}{2}B_{a\,\mu}(-\tilde{\omega},-\tilde{k}) \coth(\pi\tilde{\omega}) \left(\Pi_{-}^{\mu \nu}(\tilde{\omega},\tilde{k})-\Pi_{+}^{\mu \nu}(\tilde{\omega},\tilde{k})\right)B_{a\,\nu}(\tilde{\omega},\tilde{k})\bigg)\,,
\end{split}
\end{align}
where the summation only over $t$ and $x$ indices comes from restricting to the longitudinal channel (the only one with the gapless excitations), see eq.~\eqref{eq.longchann}. Note that given eq.~\eqref{eq:Z:onshell}, the terms mixing $t$ and $x$ come into play from subleading behavior of $B$'s at the UV boundary~$\partial {\cal M}$.
In eq.~\eqref{eq:SKaction} we have also defined the average and difference combinations of the pullback sources, see also eq.~\eqref{eq:B},
\be
B_{r}=\frac{1}{2}\left(B_{R}+B_L\right)\,,\qquad B_a=B_R-B_L\,,
\ee
and, on top, we also defined the following quantities
\be
\label{eq:Pi}
\Pi^{tt}_{\pm}=-G^{tt}_{\pm}\,,\qquad \Pi^{tx}_{\pm}=-G^{tx}_{\pm}\,,\qquad \Pi^{xt}_{\pm}=G^{xt}_{\pm}\,,\qquad \Pi^{xx}_{\pm}=G^{xx}_{\pm}\,,
\ee
with
\be 
 G_{\pm}^{\mu \nu} =\frac{\pi^2 T^2 L}{g_A^2} \partial_u\Delta^{\mu \nu}_{\pm}\big|_{u=0}\,.
\ee
Given the $\Delta^{\mu\nu}_{\pm}$'s  defined in \eqref{eq:deltas}, we explicitly have
\begin{align}
\begin{split}
\label{eq:Gpm}
&G^{tt}_{\pm}=\frac{\pi^2 T^2 L}{g_A^2}\frac{h_{\pm}^0n^\prime-q^0g_{\pm}^\prime}{h_{\pm}^0n^0-g_{\pm}^0q^0}\bigg|_{u=0}\,,\qquad G^{tx}_{\pm}=\frac{\pi^2 T^2 L}{g_A^2}\frac{n^0g_{\pm}^\prime-g_{\pm}^0n^\prime}{h_{\pm}^0n^0-g_{\pm}^0q^0}\bigg|_{u=0}\,,\\
&G^{xt}_{\pm}=\frac{\pi^2 T^2 L}{g_A^2}\frac{h_{\pm}^0q^\prime-q^0h_{\pm}^\prime}{h_{\pm}^0n^0-g_{\pm}^0q^0}\bigg|_{u=0}\,,\qquad G^{xx}_{\pm}=\frac{\pi^2 T^2 L}{g_A^2}\frac{n^0h_{\pm}^\prime-g_{\pm}^0q^\prime}{h_{\pm}^0n^0-g_{\pm}^0q^0}\bigg|_{u=0}\,,
\end{split} 
\end{align}
where ${}^\prime$ denotes a derivative with respect to the radial coordinate $u$. Note that the only dependence on the sources and gapless modes comes into eq.~\eqref{eq:SKaction} obviously through $B$'s and the functions~\eqref{eq:Gpm} depend only on $\tilde{\omega}$, $\tilde{k}$ and otherwise, trivially, on $T$. One important thing that we want to stress here is that the coefficients in front of various $B$-terms in eq.~\eqref{eq:SKaction} are not Green's functions given by eq.~\eqref{eq:Green}, since the Goldstone field has not been yet integrated out (i.e.~put on-shell). In particular, these coefficients, given by simple derivative expansions, do not have visible singularities as opposed to Green's functions, which we would expect to have a diffusive pole.

Let us now integrate out the dynamical degrees of freedom by imposing their equations of motion 
\be
\frac{\delta S_{eff}}{\delta \phi_r}=0\qquad \text{and} \qquad \frac{\delta S_{eff}}{\delta \phi_a}=0\,.
\ee
As anticipated above, the holographic Schwinger-Keldysh effective action~\eqref{eq:SKaction} put on shell becomes the generating functional for connected correlators
\begin{align}
 \begin{split}
 \label{eq:W}
W= &\int \frac{d\omega\,d\vec{k}}{(2\pi)^4}\bigg( Z_{a}(-\tilde{\omega},-\tilde{k})\,\Pi_-(\tilde{\omega},\tilde{k})\,Z_r(\tilde{\omega},\tilde{k})+Z_{r}(-\tilde{\omega},-\tilde{k})\,\Pi_+(\tilde{\omega},\tilde{k})\,Z_a(\tilde{\omega},\tilde{k})\\
 &\qquad +\frac{1}{2}Z_{a}(-\tilde{\omega},-\tilde{k})\,\coth(\pi\tilde{\omega})\left(\Pi_-(\tilde{\omega},\tilde{k})-\Pi_+(\tilde{\omega},\tilde{k})\right)\,Z_a(\tilde{\omega},\tilde{k})\bigg)\,,
\end{split}
\end{align}
with
\be
\Pi_{\pm}= \frac{G_{\pm}^{tt}\,G_{\pm}^{xx}-G_{\pm}^{tx}\,G_{\pm}^{xt}}{\tilde{\omega}^2\,G_{\pm}^{tt}-\tilde{k}\,\tilde{\omega}\,\left(G_{\pm}^{tx}-G_{\pm}^{xt}\right)-\tilde{k}^2\,G_{\pm}^{xx}}\,,
\ee
and we have defined the fully gauge-invariant variable for sources
\be\label{eq:defZ}
Z=-i\,\tilde{k}\,A_t-i\,\tilde{\omega}\,A_x\,.
\ee

By varying \eqref{eq:W} with respect to the external gauge invariant sources $Z$, the formal expressions for the corresponding retarded, advanced and symmetric two-point functions in momentum space are given by 
\begin{align}
\begin{split} 
\label{eq:greenfull}
& G_{ret}(\tilde{\omega},\tilde{k})=  
i\left(\Pi_-(\tilde{\omega},\tilde{k})+\Pi_+(-\tilde{\omega},-\tilde{k})\right)\,,\\
& G_{adv}(\tilde{\omega},\tilde{k})= 
i\left(\Pi_+(\tilde{\omega},\tilde{k})+\Pi_-(-\tilde{\omega},-\tilde{k})\right)\,,\\
& G_{sym}(\tilde{\omega},\tilde{k})= 
\frac{1}{2}\coth(\pi\tilde{\omega})\left(\Pi_-(\tilde{\omega},\tilde{k})-\Pi_+(\tilde{\omega},\tilde{k})+\Pi_+(-\tilde{\omega},-\tilde{k})-\Pi_-(-\tilde{\omega},-\tilde{k})\right)\,,\\
\end{split}
\end{align}
c.f. eq.~\eqref{eq:Green}.
It is straightforward to verify that the usual fluctuation-dissipation theorem for two-point thermal correlators is satisfied 
\be\label{eq:FDT}
i\,G_{sym}(\tilde{\omega},\tilde{k})=\frac{1}{2}\coth(\pi \tilde{\omega})\left(G_{ret}(\tilde{\omega},\tilde{k})-G_{adv}(\tilde{\omega},\tilde{k})\right)\,,
\ee
and that
\be
G_{ret}(\tilde{\omega},\tilde{k})=G_{adv}(-\tilde{\omega},-\tilde{k})\,.
\ee
The usual current-current two-point functions can now be easily recovered from the expressions \eqref{eq:greenfull} by multiplying with appropriate $\tilde{\omega}$ and $\tilde{k}$ factors.

\subsection{The Schwinger-Keldysh effective action up to second order}

While the discussion in previous section was formal, it is certainly of interest to have a look at the explicit form of the  effective action up to third  order in derivatives which gives rise to constitutive relations at second order\footnote{Remember $B_a\sim{\partial}$.}. Consider the modes \eqref{eq:modes} and their solutions.
To first order in a hydrodynamic expansion, the solutions take the form
\begin{align}
\begin{split}
h_{\pm}=(1-u)^{\pm\frac{i\tilde{\omega}}{2}}\bigg(1\pm\frac{i\tilde{\omega}}{2}{\rm ln}\left(\frac{2}{1+u}\right)\bigg)\,,\qquad n=1\,,\qquad 
g_{\pm}=0\,,\qquad 
q=0\,.
\end{split}
\end{align}
We refer interested reader to appendix \ref{A:second} for an analytic form of second order formulae. Inserting these expressions into eqs.~\eqref{eq:Gpm} and~\eqref{eq:Pi} we find 
\begin{align}
 \begin{split}
&\Pi^{tt}_{\pm} =\frac{\pi^2 T^2 L}{g_A^2}\left(1-2\tilde{k}^2\ln(2)\right)\,,\qquad \Pi^{tx}_{\pm}=\Pi^{xt}_{\pm}=-\frac{\pi^2 T^2 L}{g_A^2}\tilde{k}\tilde{\omega}\ln(2)\,,\\
& \Pi^{xx}_{\pm}=\frac{\pi^2 T^2 L}{g_A^2}\left(\pm i\,\tilde{\omega}-\tilde{\omega}^2\ln(2)\right)\,,
\end{split}
\end{align}
which, as anticipated in the previous section, are just given by truncated Taylor series in frequency and momentum, i.e. do not have any singularities at the level of their expansions.
Substituting these expressions into the effective action~\eqref{eq:SKaction} and taking the hydrodynamic limit to third order gives eq.~\eqref{eq:seffAppendix} from the appendix. Here we just quote its real space form, which is
\begin{align}
\begin{split}
\label{eq:sefffirst2}
\hspace{-8 pt}S_{eff}=&\frac{2\pi^2 T^2 L}{g_A^2}\int d^4x \,\bigg( B_{r\,t}B_{a\,t}-\frac{1}{2\pi T}B_{a\,x}\,\partial_t B_{r\,x}+\frac{i}{2\pi}B_{a\,x}B_{a\,x}\\
&+\frac{\ln(2)}{2\pi^2 T^2}B_{a\,t}\,\partial_x^2B_{r\,t}
+\frac{\ln(2)}{4\pi^2 T^2}\partial_tB_{a\,x}\,\partial_xB_{r\,t}+\frac{\ln(2)}{4\pi^2 T^2}\partial_xB_{a\,t}\,\partial_tB_{r\,x}+\frac{\ln(2)}{4\pi^2 T^2}B_{a\,x}\,\partial_t^2B_{r\,x}\bigg)\,.
\end{split}
\end{align}
Eq.~\eqref{eq:sefffirst2} is our final result for the effective action for diffusion to third order (second order constitutive relations)  for holographic field theories and to quadratic order in amplitudes.
It  obviously satisfies all the symmetries presented in sec.~\ref{S:SK} at tree level, i.e. the Schwinger-Keldysh symmetry \eqref{eq:SKsymm}, the KMS symmetry \eqref{eq:KMSsymm}, etc., as it should.

The constitutive relations that can be extracted from the effective action \eqref{eq:sefffirst2} are
\begin{align}
J^t &=\frac{2\pi^2 T^2 L}{g_A^2} \mu +\frac{\ln(2) L}{g_A^2}\partial_x^2 \mu+\frac{\ln(2)L}{2g_A^2}\partial_x V_x\,,\\
J^x &=\frac{\pi T L}{g_A^2}V_x -\frac{\ln(2)L}{2g_A^2}\partial_tE_x\,,
\end{align}
where we have identified the chemical potential $\mu=B_{r\,t}$, the electric field $E_x=\partial_x B_{r\,t}-\partial_t B_{r\,x}$ and the vector $V_x=E_x-\partial_x\mu=-\partial_tB_{r\,x}$. Although it is tantalizing to directly compare the resulting effective action and constitutive relations with the general expressions at the end of sec. \ref{S:SK}, one should bear in mind that the expressions derived here  are given in a particular reference frame. Hydrodynamic variables, like the chemical potential, can always be redefined order by order by performing a transformation  $\mu\rightarrow \mu +\delta\mu$ where $\delta \mu$ is of higher order in derivatives. In this way it is always possible to reabsorb certain transport coefficients into redefinitions of lower order ones. One possible way to extract physically independent transport coefficients is to consider frame invariant combinations of the constitutive relations, as in ref.~\cite{Bhattacharya:2011tra}. 

The alternative is to instead inspect the poles of the field theory correlators and this is the route that we will take. Up to fist order in the hydrodynamic expansion, the relevant correlators \eqref{eq:greenfull} are
\be
G_{ret}=i\,\frac{\sigma}{D\,k^2-i\,\omega}\,,\qquad G_{adv}=i\,\frac{\sigma}{D\,k^2+i\,\omega}\,,\qquad G_{sym}=i\,\frac{T\sigma}{D^2\,k^4+\omega^2}\,,
\ee
where we have identified the susceptibility and the diffusion constant as
\be
\chi=\frac{2\pi^2 T^2 L}{g_A^2}\,,\qquad D=\frac{\sigma}{\chi}=\frac{1}{2\pi T}\,.
\ee
These values match  known results derived from holography by other means, see, e.g., \cite{Policastro:2002se,Kovtun:2008kx}.

From the third order effective action \eqref{eq:sefffirst2} we can compute the corresponding  generating functional and extract, say, the diffusive pole to second order in the hydrodynamic expansion
\be
\label{eq:omdiffholo}
\omega = -\frac{i}{2\pi T}k^2\left(1+\frac{\ln(2)}{(2\pi T)^2}k^2\right)\,,
\ee
which agrees with known results in literature, see, e.g., \cite{Policastro:2002se}.
 A comparison with the field theory result \eqref{eq:diffsecond} gives us\footnote{We thank P. Kovtun for discussions on this point.} 
\be
\frac{1}{\chi}\left(k_1-2k_2+k_3-sD\right)=\frac{\ln(2)}{(2\pi T)^2}\,.
\ee

To close this section, let us re-stress that the key novelty of the presented approach are not things like eq.~\eqref{eq:omdiffholo}, which has been already known for a long time, but rather the ability to reproduce from an ab initio calculation structures postulated earlier using an effective field theory reasoning. 

\section{The near-horizon region and the Schwinger-Keldysh effective action}
\label{S:IRmaster}

In this section we derive the Schwinger-Keldysh effective action describing diffusion on an intermediate radial cut-off of the bulk spacetime, which interpolates between the horizons and the conformal boundaries. 
Fluid dynamics on finite cutoff hypersurfaces for black holes in holography and beyond was considered before in refs.~\cite{Bredberg:2010ky,Cai:2011xv,Brattan:2011my,Bai:2012ci,Emparan:2013ila,Pinzani-Fokeeva:2014cka} and here we focus on the charge dynamics to understand its role, via semi-holography~\cite{Faulkner:2010tq,Faulkner:2010jy,Nickel:2010pr}, in calculations of Schwinger-Keldysh effective actions.

\subsection{The infrared Schwinger-Keldysh effective action}
\label{S:IRlimit}

The ansatz for the gauge field in the bulk is similar to the one discussed  around eqs.~\eqref{Eq:ansatzB} and~\eqref{eq:modes}. The Dirichlet boundary conditions  are now fixed on timelike hypersurfaces at finite radius $u=\delta$ as follows
\begin{align}
\begin{split}
B^{\delta}_{s\,\mu}(t_s,x)=B_{s\,\mu}(t_s,x,u=\delta) \,,\qquad B^{\delta}_{E\,\mu}(\tau,x)=B_{E\,\mu}(\tau,x,u=\delta)\,,
\end{split} 
\end{align}
where $s=R,L$ and $\mu=t,x$ as usual. 
This time we have allowed non-vanishing sources on the (finite cutoff) boundary of the Euclidean manifold as well. 
As previously, we impose that the gauge field at the horizon is vanishing, namely we require eqs.~\eqref{eq:conconst} and~\eqref{eq:conconst2}, to secure regularity of the gauge field at the tip of the Euclidean cigar. Gluing conditions along the fixed time slices at $t=0$ and at $t=+\infty$, similarly to eqs.~\eqref{eq:condtime} and~\eqref{eq:apL},
imply 
\begin{align}
\begin{split}
\label{eq:as}
&  a_+^E=a_+^L\,,\qquad a_+^R=e^{-\omega/T}a_+^L\,, \\
& a_+^L=(1+\bar{n}(\omega))\frac{(B_{L\,x}-B_{R\,x})n^{\delta}-(B_{L\,t}-B_{R\,t})q^{\delta}}{h_{+}^{\delta}n^{\delta}-g_{+}^{\delta}q^{\delta}}\,,
\end{split} 
\end{align}
where we have defined the shortcut notation for, e.g., $h_+^{\delta}=h_+(\tilde{\omega},\tilde{k},u=\delta)$.

The solution in the Lorentzian parts of the bulk spacetime can be written similarly to eq.~\eqref{sol}~as
\begin{align}
\begin{split}
B_{s\,\mu}(t,x,u)=\int\frac{d\omega\, d\vec{k}}{(2\pi)^4}\,e^{-i\,\omega \,t_s+i\,k\,x}\sum_{m=R,L}\sum_{\nu=t,x}\Delta_{sm}^{\delta\,\,\mu\nu}(\tilde{\omega},\tilde{k},u)B_{m\,\nu}^{\delta}(\tilde{\omega},\tilde{k})\,, 
\end{split}
\end{align}
where the bulk to boundary propagators now depend on the position of the radial cutoff $\delta$,
\begin{align}
\begin{split}
& \Delta_{LL}^{\delta\,\,\mu\nu}=  -\bar{n}(\omega)\Delta_{-}^{\delta\,\,\mu\nu}+(1+\bar{n}(\omega)){\Delta}_{+}^{\delta \,\,\mu\nu}\,,\qquad 
\Delta_{LR}^{\delta \,\,\mu\nu}=  (1+\bar{n}(\omega))\left(\Delta_{-}^{\delta \,\,\mu\nu}-\Delta_{+}^{\delta \,\,\mu\nu}\right)\,,\\
& \Delta_{RL}^{\delta \,\,\mu\nu}=  -\bar{n}(\omega)\left(\Delta_{-}^{\delta \,\,\mu\nu}-\Delta_{+}^{\delta \,\,\mu\nu}\right)\,,\qquad
\Delta_{RR}^{\delta \,\,\mu\nu}=  (1+\bar{n}(\omega))\Delta_{-}^{\delta \,\,\mu\nu} -\bar{n}(\omega)\Delta_{+}^{\delta \,\,\mu\nu}\,, 
\end{split}
\end{align}
and we have defined the combinations 
\begin{align}
\begin{split}
&\Delta^{\delta \,\,tt}_{\pm}=\frac{h_{\pm}^{\delta} n-q^{\delta}g_{\pm}}{h_{\pm}^{\delta}n^{\delta}-g_{\pm}^{\delta}q^{\delta}}\,,\qquad \Delta^{\delta \,\,tx}_{\pm}=\frac{n^{\delta}g_{\pm}-g_{\pm}^{\delta}n}{h_{\pm}^{\delta}n^{\delta}-g_{\pm}^{\delta}q^{\delta}}\,,\\
&\Delta^{\delta \,\,xt}_{\pm}=\frac{h_{\pm}^{\delta}q-q^{\delta}h_{\pm}}{h_{\pm}^{\delta}n^{\delta}-g_{\pm}^{\delta}q^{\delta}}\,,\qquad \Delta^{\delta \,\,xx}_{\pm}=\frac{n^{\delta}h_{\pm}-g_{\pm}^{\delta}q}{h_{\pm}^{\delta}n^{\delta}-g_{\pm}^{\delta}q^{\delta}}\,.
\end{split} 
\end{align}
The solution in the Euclidean manifold can be written in an analogous fashion.

The infrared Schwinger-Keldysh effective action can be now derived evaluating the partially on-shell effective action in the various parts of the spacetime, as we have done around eq.~\eqref{eq:Sonshell}. The result is
\begin{align}
\begin{split}
\label{eq:SeffIR}
\hspace{-7 pt}S_{eff}^{\delta}=&\int \frac{d\omega\, d\vec{k}}{(2\pi)^4}\sum_{\mu=t,x}\sum_{\nu=t,x}\bigg(
B^{\delta}_{a\,\mu}(-\tilde{\omega},-\tilde{k})\Pi_{-}^{\delta\,\,\mu\nu}(\tilde{\omega},\tilde{k})B^{\delta}_{r\,\nu}(\tilde{\omega},\tilde{k})+B^{\delta}_{r\,\mu}(-\tilde{\omega},-\tilde{k})\Pi_{+}^{\delta\,\,\mu\nu}(\tilde{\omega},\tilde{k})B^{\delta}_{a\,\nu}(\tilde{\omega},\tilde{k})\\
&\qquad +\frac{1}{2}B^{\delta}_{a\,\mu}(-\tilde{\omega},-\tilde{k}) \coth(\pi\tilde{\omega}) \left(\Pi_{-}^{\delta\,\,\mu\nu}(\tilde{\omega},\tilde{k})-\Pi_{+}^{\delta\,\,\mu\nu}(\tilde{\omega},\tilde{k})\right)B^{\delta}_{a\,\nu}(\tilde{\omega},\tilde{k})\bigg)\,,
\end{split}
\end{align}
where we have defined
\be
\Pi^{\delta\,\,tt}_{\pm}=-G^{\delta\,\,tt}_{\pm}\,,\qquad \Pi^{\delta\,\,tx}_{\pm}=-G^{\delta\,\,tx}_{\pm}\,,\qquad \Pi^{\delta\,\,xt}_{\pm}=(1-\delta^2)G^{\delta\,\,xt}_{\pm}\,,\qquad \Pi^{\delta\,\,xx}_{\pm}=(1-\delta^2)G^{\delta\,\,xx}_{\pm}\,,
\ee
and
\be 
 G_{\pm}^{\delta\,\,\mu\nu} =\frac{\pi^2 T^2 L}{g_A^2} \partial_u\Delta^{\delta\,\,\mu\nu}_{\pm}\big|_{u=\delta}\,.
\ee
In writing eq.~\eqref{eq:SeffIR} we have omitted the contribution of the Euclidean cap for simplicity which is not vanishing in this case. Notice that by sending the intermediate cut-off to the conformal boundary $\delta\rightarrow 0$ the original Schwinger-Keldysh effective action \eqref{eq:SKaction} is recovered.
Moreover, we can clearly see that the same characteristics of the full effective action of the dual field theory are retained in the infrared. For instance, the effective action is at least linear in the $a$-type sources to satisfy the Schwinger-Keldysh symmetry of the underlying path integral.

Let us now consider the limit where the intermediate radial cutoff is taken to be very close to the horizon  $\delta\rightarrow 1$. The Schwinger-Keldysh effective action \eqref{eq:SeffIR} becomes
\begin{align}
\begin{split}
\label{eq:seffIR2}
S_{eff}^{IR}=&\frac{\pi^2 T^2 L}{g_A^2}\int \frac{d\omega\, d\vec{k}}{(2\pi)^4}\,\bigg( \frac{1}{(1-\delta)}\left(B^{\delta}_{r\,t}(-\tilde{\omega},-\tilde{k})B^{\delta}_{a\,t}(\tilde{\omega},\tilde{k}) +B^{\delta}_{r\,t}(\tilde{\omega},\tilde{k})B^{\delta}_{a\,t}(-\tilde{\omega},-\tilde{k})\right)\\
&\qquad  -i\,\tilde{\omega}B^{\delta}_{r\,x}(-\tilde{\omega},-\tilde{k})B^{\delta}_{a\,x}(\tilde{\omega},\tilde{k})+i\,\tilde{\omega}B^{\delta}_{r\,x}(\tilde{\omega},\tilde{k})B^{\delta}_{a\,x}(-\tilde{\omega},-\tilde{k})\\
&\qquad +i\,\tilde{\omega}\coth(\pi\tilde{\omega})B^{\delta}_{a\,x}(-\tilde{\omega},-\tilde{k})B^{\delta}_{a\,x}(\tilde{\omega},\tilde{k})\bigg)\,.
\end{split}
\end{align}
This effective action is universal and simple in the sense that it is given to all orders in  a hydrodynamic expansion since in deriving it we have not used the explicit mode solutions but only their near horizon behaviors. In this way this object can be thought of as a simple effective action incorporating already all the symmetries of the full Schwinger-Keldysh effective action even if it is restricted to some near horizon region. 

To make comparison with known results  we  need to rescale the coordinates in a way that  the induced metric on the intermediate radial cutoff $\delta$ is (conformally) flat
\be
ds^2 = \frac{(\pi T L)^2}{u}\left(-f(u)dt^2+dx^2+dy^2+dz^2\right)=  \frac{(\pi T L)^2}{u}\left(-d\bar{t}^2+dx^2+dy^2+dz^2\right)\,,
\ee
with  $\bar{t}=\sqrt{f_{\delta}}\,t$ and $f_{\delta} = (1-\delta^2)$. With this redefinition the various field theory quantities scale as 
\be
\bar{\omega}=\omega/\sqrt{f_{\delta}}\,,\qquad \bar{T}=T/\sqrt{f_{\delta}}\,,\qquad B_{\bar{t}}=B_t/\sqrt{f_{\delta}}\,,
\ee
and the near horizon effective action becomes 
\begin{align}
\begin{split}
\label{eq:seffIR4}
\bar{S}_{eff}^{IR}=&\frac{\pi^2\,f_{\delta}\,\bar{T}^2 L}{g_A^2}\int \frac{d\bar{\omega}\, d\vec{k}}{(2\pi)^4}\sqrt{f_{\delta}}\,\bigg(2\left(B^{\delta}_{r\,\bar{t}}(-\tilde{\omega},-\tilde{k})B^{\delta}_{a\,\bar{t}}(\tilde{\omega},\tilde{k}) +B^{\delta}_{r\,\bar{t}}(\tilde{\omega},\tilde{k})B^{\delta}_{a\,\bar{t}}(-\tilde{\omega},-\tilde{k})\right)\\
& -i\,\frac{\bar{\omega}}{2\pi \bar{T}}B^{\delta}_{r\,x}(-\tilde{\omega},-\tilde{k})B^{\delta}_{a\,x}(\tilde{\omega},\tilde{k})+i\,\frac{\bar{\omega}}{2\pi\bar{T}}B^{\delta}_{r\,x}(\tilde{\omega},\tilde{k})B^{\delta}_{a\,x}(-\tilde{\omega},-\tilde{k})\\
&+i\,\frac{\bar{\omega}}{2\pi \bar{T}}\coth\left(\frac{\bar{\omega}}{2 \bar{T}}\right)B^{\delta}_{a\,x}(-\tilde{\omega},-\tilde{k})B^{\delta}_{a\,x}(\tilde{\omega},\tilde{k})\bigg)\,,
\end{split}
\end{align}
which is effectively living  on the worldvolume.
The diffusion constant in these coordinates is  given by
\be
D=\frac{\sigma}{\chi}=\frac{1}{4\pi \bar{T}}
\ee
and matches the well-known result of ref.~\cite{Bredberg:2010ky} with an \emph{exact} dispersion relation $\bar{\omega}=-i\,D\,k^2$. The diffusion constant on a finite radial cutoff can also be easily extracted $D=1/2\pi \bar{T} (1+\delta)$ which interpolates between the field theory result defined on the conformal  boundary and the one on the horizon, see, again,~ref.~\cite{Bredberg:2010ky}.

\subsection{The infrared/ultraviolet coupling} 

Having the infrared Schwinger-Keldysh effective action \eqref{eq:seffIR2}  as a simple object incorporating the near horizon physics we would like to show that  the remaining part of the spacetime between the horizon and the conformal boundary acts as a mere transfer of this information to the dual conformal field theory.

To see this, let us use the finite radial cutoff to separate the bulk spacetime into an IR region ranging between the cutoff and the horizon  and a UV region extending from the cutoff to the conformal boundary, see, e.g., \cite{Nickel:2010pr,Heemskerk:2010hk,Faulkner:2010jy}. The names associated to the different regions in spacetime parallel the field theory interpretation where, through the UV/IR relation in holography, short/large scale physics in the bulk corresponds to the IR/UV regime in the dual field theory. With this division of the bulk spacetime, also the bulk action separates artificially  into an IR and an UV contribution 
\be
\label{eq:Stot}
S = S^{IR}+S^{UV}\,.
\ee
We have to imagine that the intermediate timelike cutoff hypersurface acts as a boundary to the IR spacetime and as a second boundary to the UV spacetime on which we impose Dirichlet boundary conditions $B^{\delta}$. The full action is then recovered when imposing continuity conditions along the cutoff hypersurface
\be
\label{eq:continuity}
\frac{\delta S^{IR}}{\delta B^{\delta}}=\frac{\delta S^{UV}}{\delta B^{\delta}}\,.
\ee

The IR effective action   where the cutoff is taken very close to the horizon is given by eq.~\eqref{eq:seffIR2}. The question is whether it is indeed sufficient to recover the full Schwinger-Keldysh effective action \eqref{eq:SKaction} once also the UV part of the effective action is known.
The UV effective action can be computed considering a double-Dirichlet problem along the lines of what we did in ref.~\cite{deBoer:2015ija} for linearized gravity perturbations, see also ref.~\cite{Crossley:2015tka}. For instance, we first solve the dynamical equations of motion \eqref{eq:Bt} and \eqref{eq:Bx} for the bulk gauge field subject to two Dirichlet boundary conditions, $B^{\delta}$ at the intermediate cutoff $u=\delta$  and $B$ at the conformal boundary $u=0$. Up to leading order in a hydrodynamic expansion the solution is simply given by
\begin{align}
\begin{split}
\label{eq:solDD}
& B_t(\tilde{\omega},\tilde{k},u)=\left(1-\frac{u}{\delta}\right)B_t+\frac{u}{\delta}B_t^{\delta}\,,\\
& B_x(\tilde{\omega},\tilde{k},u)=\left(1-\frac{\text{ArcTanh}(u)}{\text{ArcTanh}(\delta)}\right)B_x+\frac{\text{ArcTanh}(u)}{\text{ArcTanh}(\delta)}B_x^{\delta}\,.
\end{split}
\end{align}
Subsequently, we evaluate the partially on-shell bulk action \eqref{eq:Z:onshell} on the solution \eqref{eq:solDD} where, this time,  the boundary $\partial{\cal M}$ is taken to be the intermediate cutoff and the conformal boundary. Thus, taking three such combinations
\be
iS_{eff}^{UV} = iS_{\rm onshell}\bigg|_{u_R=0}^{u_R=\delta}-iS_{\rm onshell}\bigg|_{u_L=0}^{u_L=\delta}-S_{\rm onshell}\bigg|_{u_E=0}^{u_E=\delta}
\ee
 gives us the total effective action in the UV region of our mixed signature spacetime
\begin{align}
\begin{split}
\label{eq:effUV}
&S_{eff}^{UV} = \frac{\pi^2 T^2 L}{g_A^2}\int \frac{d\omega\, d\vec{k}}{(2\pi)^4}\,\bigg(\frac{1}{\delta}\left(B_{R\,t}(-\tilde{\omega},-\tilde{k})-B_{R\,t}^{\delta}(-\tilde{\omega},-\tilde{k})\right)\left(B_{R\,t}(\tilde{\omega},\tilde{k})-B_{R\,t}^{\delta}(\tilde{\omega},\tilde{k})\right)\\
&-\frac{1}{\text{ArcTanh}(\delta)}\left(B_{R\,x}(-\tilde{\omega},-\tilde{k})-B_{R\,x}^{\delta}(-\tilde{\omega},-\tilde{k})\right)\left(B_{R\,x}(\tilde{\omega},\tilde{k})-B_{R\,x}^{\delta}(\tilde{\omega},\tilde{k})\right)-(R \rightarrow L)\bigg)\,,
\end{split}
\end{align}
where we have neglected the Euclidean contribution for simplicity.
 Gluing conditions along fixed time slices can be easily satisfied requiring that the various Dirichlet values are equal at the crossing points.
 
Notice that taking strictly a near horizon limit of $S_{eff}^{UV}$ in eq.~\eqref{eq:effUV} gives a  very simple effective action which only depends on the time component of the gauge field $B_t$. However, given that the IR effective action $S_{eff}^{IR}$ in eq.~\eqref{eq:seffIR2} is divergent in this limit, we will posit to stay slightly away from the horizon. In this way we can safely impose the continuity condition~\eqref{eq:continuity} among the IR and UV effective actions. Integrating out the Dirichlet values at the finite cutoff, taking the hydrodynamic limit first and the near horizon limit second, we recover the Schwinger-Keldysh effective action up to second order in a derivative expansion~\eqref{eq:sefffirst2}. We have also verified that the same holds  to third order  when the UV effective action $S_{eff}^{UV}$ is given to second order while keeping the universal IR effective action $S_{eff}^{IR}$ in eq.~\eqref{eq:seffIR2}. This strongly suggest the picture advocated at the beginning of this subsection, that is the near horizon part of the spacetime can be approximated by a simple term~\eqref{eq:seffIR2} which alone includes the dissipative effects, KMS conditions etc, while the UV part of the spacetime pushes that information to the conformal boundary where the dual field theory is defined.

\subsection{Interpretation of the infrared theory}
\label{S:IRtheory}

As we just stated, the IR effective action in eq.~(\ref{eq:seffIR2}) is believed to be solely responsible for 
most of the interesting physics that
underlies the full effective action. It is therefore worthwhile to try to provide a different perspective on this IR theory.
In general, whenever we have a black hole or black brane with a horizon at some value $r=r_S$ of a radial coordinate~$r$,
we write $r=r_S +\epsilon\,\rho$ and consider the limit $\epsilon \rightarrow 0$ in order to take a near horizon limit. In this
limit the metric behaves schematically as 
\begin{equation} \label{nhmetric}
ds^2 \sim \epsilon \,(d\rho^2 - \rho^2 d{\cal T}^2) + d\Omega_H + \ldots
\end{equation} 
where $\rho$ and ${\cal T}$ are two-dimensional Rindler coordinates and $\Omega_H$ is the metric on the horizon. 

We first observe that time derivatives with respect to~${\cal T}$ will come with an extra factor of $1/\sqrt{\epsilon}$ compared
to spatial derivatives along the horizon, and as a result the former will dominate. This is why in the IR effective
action the momenta $k$ do not appear explicitly and the action is ultra-local along the horizon. It also explains why
the IR effective action can be obtained from a Kaluza-Klein reduction of the full theory along the horizon directions.
For the case studied in this paper, a higher dimensional gauge field gives rise to some scalar fields and one gauge field
in two dimensions. More precisely, $B_t$ corresponds to a two-dimensional gauge field, and $B_x$ to a two-dimensional scalar. In two dimensions these
fields do not mix and indeed, in the IR effective action they do not mix.

The two-dimensional action for scalars is conformally invariant and does therefore not explicitly depend on the parameter $\epsilon$.
Since $\epsilon\sim (1-\delta)$, this explains the absence of a factor of $1-\delta$ in the part of the effective action that depends on $B_x$. The two-dimensional action for gauge fields on the other hand will scale as $1/\epsilon\sim 1/(1-\delta)$, which
explains the corresponding factor in the terms of the IR effective action that contains~$B_t$.

Besides all this, there is also an emergent conformal symmetry in the near-horizon limit. Rindler spacetime is just a wedge
of two-dimensional Minkowski spacetime and inherits its conformal symmetries. We are considering the effective action with some cutoff
$\rho_{c}={\rm const}$, but, as is familiar from boundary states in conformal field theory, this boundary will preserve half
of the conformal symmetries. In terms of light-cone coordinates $x^{\pm}=\rho\,e^{\pm {\cal T}}$, if we map $x^+ \rightarrow g(x^+)$
and $x^-\rightarrow \rho_c^2/g(\rho_c^2/x_-)$ for any function~$g$ then the boundary $x^+ x^-=\rho_c^2$ will be mapped to itself. 

If we interpret the IR effective action as a quantum mechanical effective action for each value of the momenta $k$,
with only $\omega$-dependence, then this effective action will inherit the above conformal invariance (at least
for the scalar degrees of freedom). In other words, 
we can reinterpret part of eq.~(\ref{eq:seffIR2}) as the effective action of finite temperature conformally invariant quantum
mechanics. 

Indeed, if we look at $B_x$, this corresponds to a massless scalar in two-dimensions which has scaling dimension
zero. At zero temperature, the retarded and advanced two-point functions would presumably just be theta-functions
of the time difference, and the Wightman two-point function would most likely be as in two-dimensions, i.e. 
be of the form $\sim \log|{\cal T}_1-{\cal T}_2|$. We can then obtain the finite temperature two-point functions in conformal
quantum mechanics by a replacement of the form ${\cal T}_1-{\cal T}_2\rightarrow a\sinh \left[({\cal T}_1-{\cal T}_2)/a\right]$. Indeed, taking such two-point 
functions and performing a Fourier transformation leads to an effective action which is of the same form as that
given for $B_x$ in eq.~(\ref{eq:seffIR2}).

The action for $B_t$ is given by two-dimensional Yang-Mills theory which is not conformally invariant. 
The above consideration therefore does not immediately yield a useful constraint on the terms in the effective action
that contain~$B_t$. The Yang-Mills action on two-dimensional Rindler spacetime
does however have a symmetry
under ${\cal T}\rightarrow \lambda \, {\cal T}$, $A_\rho \rightarrow \sqrt{\lambda} \,A_\rho$ and $A_{\cal T} \rightarrow A_{\cal T}/\sqrt{\lambda}$ which
implies that the effective action for $B_t$ should behave under scale transformation as if $B_t$ was an operator
with scaling dimension $1/2$. This is indeed consistent with the IR effective action.

We postpone a detailed study of this IR theory and the origin of the emergent conformally invariant quantum mechanical system to future work.

\section{Discussion}
\label{S:Discussion}

Motivated by recent developments in Schwinger-Keldysh effective field theories for hydrodynamics we have explicitly derived the low-energy local effective action capturing the dynamics of a conserved $U(1)$ charge current in strongly coupled conformal quantum field theories at large-$N$ using holographic techniques. Our results  match existing literature both in  field theory as far as the formal structure of the effective action is concerned and in holography for the actual values of the transport coefficients.    

One main virtue of our approach, which utilizes real-time holography techniques developed in refs.~\cite{Skenderis:2008dh,Skenderis:2008dg}, is that it allows  to obtain analyticity conditions on the thermal correlators (KMS conditions) without relying on particular analytic continuations of fields across horizons. The KMS conditions arise naturally in this holographic setup as a result of the gluing conditions of the Lorentzian spacetimes with the Euclidean cap along fixed initial and final time slices, see also ref.~\cite{Botta-Cantcheff:2018brv}.

Our derivation so far has been limited to the simplest case of a probe gauge field without backreaction on the bulk metric and, as such, it is not capturing the  dynamics of a conserved stress-energy tensor. One obvious obstruction to achieve a fully-fledged fluid dynamical behavior is the need for a more careful treatment of the matching conditions at the fixed time slices which now would  be fluctuating. We expect our construction to generalize at least to linearized order in the gravitational field perturbations, capturing the linearized fluid dynamic effective action for the dual charged conformal fluids at strong coupling. Such construction would need a full understanding of the gauge field dynamics on top of a Reisner-Nordstrom background with backreaction. Looking ahead, deriving the nonlinear Schwinger-Keldysh effective action for fluid dynamics from gravity would require further refinements of the procedure presented in this paper. Such a result could be viewed as an alternative derivation of  the fluid/gravity correspondence~\cite{Bhattacharyya:2008jc}.

One feature that we have not seen in our construction is a geometrical
interpretation of the ghost fields which we have completely neglected in this work. This is
not so surprising since we have only restricted ourselves to the tree level effective actions
in the~$1/N$~expansion. The ghost fields appear in order to interpret the vanishing of the
effective action for particular source configurations as a consequence of certain BRST
symmetries, see refs.~\cite{Haehl:2015foa,Crossley:2015evo,Haehl:2015uoc,Haehl:2016pec,Haehl:2016uah}, see also refs.~\cite{Haehl:2015uoc,Haehl:2016pec,Haehl:2016uah,Jensen:2017kzi,Gao:2017bqf,Haehl:2018lcu,Jensen:2018hse} for a supersymmetric implementation of these symmetries. As we saw, the relevant source configurations have a very simple IR interpretation, as
they correspond to purely left-moving and purely right-moving field configurations in the near-horizon
Rindler region. This suggests that, by gauge fixing the 2-dimensional diffeomorphisms of the IR Rindler geometry, 
the relevant ghosts may naturally appear. This is quite familiar from the standard treatment of the
string world-sheet, where one also gauge fixes the 2-dimensional diffeomorphisms and as a result the left- and
right-moving stress tensor become BRST exact. We hope to report on this direction in more detail in the future and to see how the results parallel  the discussion in ref.~\cite{Gao:2018bxz}.

Another important aspect that we have not  addressed in this work is the entropy current. One of the major results in Schwinger-Keldysh effective field theories for hydrodynamics is the possibility to define the entropy current as a Noether current of a symmetry of the Schwinger-Keldysh effective action \cite{Glorioso:2016gsa}, see also \cite{Jensen:2018hhx,Haehl:2018uqv,Jensen:2018hse} for a supersymmetric implementation. Such a symmetry should have a dual version in the bulk and we postpone this derivation to future work.

It is important to notice that our results obtained from holography are given at finite $\hbar$. An effective action and an entropy current for diffusion using Schiwnger-Keldysh techniques at finite $\hbar$  have been constructed in ref.~\cite{Jensen:2017kzi,Jensen:2018hhx}. Extending these results to the full fluid dynamics case is problematic and one needs to rely on an approximation where quantum fluctuations are neglected. This can be achieved by going to the so-called statistical mechanics limit where $\hbar\rightarrow 0$, see refs.~\cite{Crossley:2015evo,Glorioso:2017fpd,Jensen:2018hse}, or, equivalently, to a high-temperature regime, see refs.~\cite{Haehl:2015foa,Haehl:2015uoc,Haehl:2018lcu}. In holographic theories, however, such a limit cannot be performed while keeping a well defined hydrodynamic expansion. The mean free path in holographic field theories is $l_{mfp}\sim \hbar/T$, therefore there is no separation of scales between the gradient expansion and quantum fluctuations. 
   The current work and future developments to include full fluid dynamics might provide an invaluable testing ground to understand how to go beyond the statistical limit in constructing Schwinger-Keldysh effective field theories for  hydrodynamics which is currently an open problem.

In this work, we have also been able to show that all the nice properties of the field theory Schwinger-Keldysh effective action reside in the near horizon region of the dual black hole spacetime. We have done so by defining a simple, universal, near horizon Schwinger-Keldysh-like effective action valid to all orders in a hydrodynamic expansion which has the same symmetry structure as the original one (Schwinger-Keldysh symmetry, KMS symmetry, etc.). We have then coupled this object to the remaining bulk spacetime extending between the horizon and the conformal boundaries and showed that the full Schwinger-Keldysh effective action for the probe gauge field is recovered to second order in a hydrodynamic expansion. 
This procedure can be thought of as a generalization of the membrane paradigm viewed as a simple boundary condition, see, e.g., ref.~\cite{Iqbal:2008by,deBoer:2014xja}, to a membrane paradigm as an action principle given in the IR~part of the spacetime.
A natural generalization of this framework is to revisit the IR effective action and its relation to a membrane paradigm
in the presence of dynamical gravity. The general scaling arguments given in subsection \ref{S:IRtheory} can perhaps be
applied to gravity as well. Since a higher dimensional metric yields some scalars, gauge fields, and a 2-dimensional metric
in two-dimensions, the main new ingredient is to understand the appropriate boundary conditions for the
fluctuations of the 2-dimensional metric.

The division of the bulk spacetime between a simple infrared near horizon region and the
remaining ultraviolet spacetime is a precise implementation of the idea of semi-holography~\cite{Faulkner:2010tq}. 
It is then natural to ask what are the different possible near horizon Schwinger-Keldysh-like actions 
when the near horizon region itself is different from the simple case analyzed in this work. For instance, what
would it be for near-extremal black holes? Near extremal black holes develop a long AdS$_2$ throat
region, and one might try to separate the bulk spacetime in an IR region which contains the
AdS$_2$ throat and the rest. In this case, the IR region can perhaps be well approximated by 
the Schwarzian theory describing nearly AdS$_2$ physics, see, e.g., ref.~\cite{Maldacena:2016upp}. We cannot, however, go arbitrarily close to the horizon, as this would not preserve the nearly AdS$_2$ 
property of the metric. We therefore do not expect a precise membrane paradigm-like description similar to 
the one for Rindler space, and that in order to reproduce the complete physical picture higher order
corrections to the IR theory need to be kept. This is in line with the observations in, e.g., refs.~\cite{Moitra:2018jqs} 
and~\cite{Castro:2018ffi}.

Another natural extension of our work is to dual field theory contours defined on multiple time segments. The effective field theories arising in these scenarios would include out-of time ordered correlators, observables which acquired much recent attention as probes of early-time chaotic behavior  leading to the celebrated result that black holes are the maximally chaotic systems in nature~\cite{Maldacena:2015waa}.   
More recently, much effort has been put into framing quantum chaos into  an effective field theory language~\cite{Blake:2017ris}, see also refs.~\cite{Haehl:2018izb,Cotler:2018zff} for the case of underlying two-dimensional conformal field theories. Our techniques can be used to test these developments via simple holographic derivations and possibly guide to a better understanding of universal features of quantum chaos.

Finally, it would be very interesting to use our construction to understand the large-order behaviour of hydrodynamic gradient expansion of the Schwinger-Keldysh effective action. What is well-established by now is that for two classes of non-linear solutions of relativistic hydrodynamics the energy-momentum tensor expanded in gradients receives numerical contributions scaling like $n!$ from the combined terms having $n$ derivatives of fluid variables, see ref.~\cite{Florkowski:2017olj} for a review. What was also seen, see refs.~\cite{Heller:2016gbp} and~\cite{Withers:2018srf}, is that hydrodynamic dispersion relations like eq.~\eqref{eq:omdiffholo} exhibit finite radius of convergence of the small-$k$ expansion. In both cases, one can link the zero and finite radia of convergence to the presence of fast-decaying excitations not captured directly by hydrodynamics in the gradient expansion. In holography, these short-lived excitations are transient quasinormal modes of black branes. Exploring what happens at the level of an effective action is an interesting open problem that might lead to a new chapter in the studies of hydrodynamics at large orders, as well as understanding effective actions beyond the derivative expansion.

\section*{Acknowledgements}
We would like to thank N. Iqbal, K. Jensen, P. Kovtun, S. Minwalla, B. van Rees, K. Skenderis and A. Yarom for discussions. The work of NPF  is supported  in part by the National Science Foundation of Belgium (FWO) grant G.001.12 Odysseus and by the European Research Council grant no. ERC-2013-CoG 616732 HoloQosmos. NPF would also like to acknowledge Technion and University of Haifa where part of this work was done. This group was supported in part by the Israel Science Foundation under grant 504/13 and in part at the Technion by a fellowship from the Lady Davis Foundation. MPH and the Gravity, Quantum Fields and Information group at AEI are supported
by the Alexander von Humboldt Foundation and the Federal Ministry for Education and Research through the Sofja Kovalevskaja Award. NPF would also like to aknowledge KITP for hospitality during the  final stage of this project. 

\begin{appendix}


\section{Quadratic effective action for diffusion}
\label{A:first}

Let us here justify in more detail the expression~\eqref{eq:Seff} from the main text for the effective action for diffusion up to third order\footnote{The resulting constitutive relations are given up to second order.} in the derivative expansion.

The most general effective action for diffusion to all orders in a hydrodynamic expansion has been given in \cite{Crossley:2015evo}.
The expression  at finite $\hbar$ and to quadratic order in the fields was given in eq.~(4.53) of ref. \cite{Jensen:2017kzi}. Expanding that expression to third order in gradients,  going back to coordinate space, and keeping only the longitudinal sector, where $\vec{k}=(k,0,0)$, gives 
\begin{align}
\begin{split}
\label{eq:Seffcomplete}
S_{eff}&=\int d^dx\Big( \chi\, B_{r\,t}B_{a\,t}-\sigma\,   B_{a\,x}\partial_{t}B_{r\,x}+i\,\sigma\, T\,   B_{a\,x}B_{a\,x}-s\,   B_{a\,t}\partial_{t}B_{r\,t}+i\,s\, T\,  B_{a\,t}B_{a\,t}\\
&+k_0\,B_{a\,t}\partial_t^2B_{r\,t}+k_1\,B_{a\,t}\partial_x^2B_{r\,t}+k_2\,\left(\partial_xB_{a\,t}\partial_t B_{r\,x}+\partial_tB_{a\,x}\partial_xB_{r\,t}\right)+k_3\,B_{a\,x}\partial^2_tB_{r\,x}\Big)\,. 
\end{split}
\end{align}
The parameters here are related to the ones in ref.~\cite{Jensen:2017kzi} via 
\begin{align}
\begin{split}
& \chi=2\,F_{00}^{tt}\,,\qquad \sigma =-\frac{1}{T} \sigma_{00}^{xx}\,,\qquad s=-\frac{1}{T} \sigma_{00}^{tt}\,,\\ 
& k_0=2\,F_{20}^{tt}-\frac{1}{12 \,T^2}F_{00}^{tt}+\frac{1}{T}\sigma_{10}^{tt}\,,\qquad k_1=2\,F_{02}^{tt}\,,\\ 
&k_2=F_{11}^{tx}+F_{11}^{xt}\,,\qquad k_3=2\,F_{20}^{xx}+\frac{1}{T}\sigma_{10}^{xx}\,,
\end{split}
\end{align}
where we have set ${\cal R}=1$ (see the original reference) and we have used the expansions  
\begin{align}
 \begin{split}
F^{\mu\nu}(i\,\omega,i\,k)&=F^{\mu\nu}_{00}+k\omega F_{11}^{\mu\nu}-\omega^2 F^{\mu\nu}_{20}-k^2 F^{\mu\nu}_{02}+{\cal O}(\omega^3,k^3)\,,\\
\sigma^{\mu\nu}(i\,\omega,i\,k)&=\sigma^{\mu\nu}_{00}+i\omega\sigma^{\mu\nu}_{10}+{\cal O}(\omega^2,k^2)\,,\\
{\cal A}(b\, \omega) &= 1+\frac{\omega^2}{12T^2}+{\cal O}(\omega^4)\,,
\end{split}
\end{align}
Here we used $b=1/T$ and the spacetime indices $\mu,\,\nu$ stand for $t,\,x$.
Notice that, for example, there is no contribution  of the form $F_{00}^{xx}$ which would give rise to a term proportional to $B_{r\,x}B_{a\,x}$ in the effective action. Such a term would not be invariant under eq.~\eqref{cond5}. Notice also that there are terms appearing at third order which come from lower order terms due to finite $\hbar$ effects. An example of such a term is the contribution $F_{00}^{tt}$ to $k_0$.

\section{Details of the holographic results}
\label{A:second}

The mode functions at second order in the hydrodynamic expansion are
\begin{align}
\begin{split}
n= &1+\frac{\tilde{k}^2}{1-u}\left(u\,{\rm ln}(u)+(1+u)\,{\rm ln}\left(\frac{2}{1+u}\right) \right)\,,\\
g_{\pm}=&+\frac{\tilde{k}\,\tilde{\omega}}{2}\left( (1-u) +u\,{\rm ln}\left(\frac{2}{1-u^2}\right)+{\rm ln}\left(\frac{2(1-u)}{1+u}\right)+2u\,{\rm ln}(u)\right)\,,\\
h_{\pm}=&(1-u)^{\pm\frac{i\tilde{\omega}}{2}}\bigg(1\pm\frac{i\tilde{\omega}}{2}{\rm ln}\left(\frac{2}{1+u}\right)+\frac{\tilde{\omega}^2}{24}\bigg(\pi^2-9{\rm ln}^2(2)+12{\rm ln}(2){\rm ln}(1-u)\\
&-6{\rm ln}(2){\rm ln}(1+u)-12{\rm ln}\left(\frac{1-u}{2}\right){\rm ln}(1+u)-12{\rm ln}(u){\rm ln}(1+u)+3{\rm ln}^2(1+u)\\
& -12 {\rm Li}_2(1-u)-12{\rm Li}_2(-u)-12{\rm Li}_2\left(\frac{1+u}{2}\right)\bigg)\bigg)\,,\\
q=&\frac{\tilde{k}\,\tilde{\omega}}{24}\bigg(\pi^2-6{\rm ln}^2(2)+12 {\rm ln}(2){\rm ln}(1-u)-12{\rm ln}(1-u){\rm ln}(1+u)-12 {\rm ln}(u){\rm ln}(1+u)\nn\\
&+6{\rm ln}^2(1+u)-12 {\rm Li}_2(1-u)-12{\rm Li}_2(-u)-12{\rm Li}_2\left(\frac{1+u}{2}\right)\bigg)\,,
\end{split}
\end{align}
where ${\rm Li}_2$ is the Polylogarithm function. 
Inserting these expressions into eq.~\eqref{eq:Gpm} and then into eq.~\eqref{eq:Pi} we get 
\begin{align}
 \begin{split}
&\Pi^{tt}_{\pm} =\frac{\pi^2 T^2 L}{g_A^2}\left(1-2\tilde{k}^2\ln(2)\right)\,,\qquad \Pi^{tx}_{\pm}=\Pi^{xt}_{\pm}=-\frac{\pi^2 T^2 L}{g_A^2}\tilde{k}\tilde{\omega}\ln(2)\,,\\
& \Pi^{xx}_{\pm}=\frac{\pi^2 T^2 L}{g_A^2}\left(\pm i\,\tilde{\omega}-\tilde{\omega}^2\ln(2)\right)\,,
\end{split}
\end{align}
and, consequently, the Schwinger-Keldysh effective action up to second order in the~Fourier space takes the form
\begin{align}
\begin{split}
\label{eq:seffAppendix}
S_{eff}=&\frac{\pi^2 T^2 L}{g_A^2}\int \frac{d\omega\, d\vec{k}}{(2\pi)^4}\,\bigg( B_{r\,t}(-\tilde{\omega},-\tilde{k})B_{a\,t}(\tilde{\omega},\tilde{k}) +B_{r\,t}(\tilde{\omega},\tilde{k})B_{a\,t}(-\tilde{\omega},-\tilde{k})\\
& -i\,\tilde{\omega}B_{r\,x}(-\tilde{\omega},-\tilde{k})B_{a\,x}(\tilde{\omega},\tilde{k})+i\,\tilde{\omega}B_{r\,x}(\tilde{\omega},\tilde{k})B_{a\,x}(-\tilde{\omega},-\tilde{k})\\
&+\frac{i}{\pi}B_{a\,x}(-\tilde{\omega},-\tilde{k})B_{a\,x}(\tilde{\omega},\tilde{k})\\
&-2\tilde{k}^2\,{\rm ln}(2)\left(B_{a\,t}(\tilde{\omega},\tilde{k})B_{r\,t}(-\tilde{\omega},-\tilde{k})+B_{a\,t}(-\tilde{\omega},-\tilde{k})B_{r\,t}(\tilde{\omega},\tilde{k})\right)\\
&-\tilde{k}\,\tilde{\omega}\,{\rm ln}(2)\left(B_{a\,x}(\tilde{\omega},\tilde{k})B_{r\,t}(-\tilde{\omega},-\tilde{k})+B_{a\,x}(-\tilde{\omega},-\tilde{k})B_{r\,t}(\tilde{\omega},\tilde{k})\right)\\
&-\tilde{k}\,\tilde{\omega}\,{\rm ln}(2)\left(B_{a\,t}(\tilde{\omega},\tilde{k})B_{r\,x}(-\tilde{\omega},-\tilde{k})+B_{a\,t}(-\tilde{\omega},-\tilde{k})B_{r\,x}(\tilde{\omega},\tilde{k})\right)\\
&-\tilde{\omega}^2\,{\rm ln}(2)\left(B_{a\,x}(\tilde{\omega},\tilde{k})B_{r\,x}(-\tilde{\omega},-\tilde{k})+B_{a\,x}(-\tilde{\omega},-\tilde{k})B_{r\,x}(\tilde{\omega},\tilde{k})\right)\bigg)\,,
\end{split}
\end{align}
where we have used a counterterm of the form for curing the divergences on the right boundary
\be
S_{ct}=\frac{L\pi^2T^2}{g_A^2}\log(u)\int \frac{d\omega\, d\vec{k}}{(2\pi)^4}\left(\tilde{k}\,B_{R\,t}(\tilde{\omega},\tilde{k})+\tilde{\omega}\,B_{R\,x}(\tilde{\omega},\tilde{k})\right)\left(\tilde{k}\,B_{R\,t}(-\tilde{\omega},-\tilde{k})+\tilde{\omega}\,B_{R\,x}(-\tilde{\omega},-\tilde{k})\right)\,,
\ee
and an analogous term for the left boundary.

\end{appendix}

\bibliographystyle{utphys}
\bibliography{2sided.bib}

\end{document}